\documentclass[prb,twocolumn,showpacs,amsmath,amssymb,letterpaper,eqsecnum,floatfix]{revtex4}

\usepackage{color,graphicx}
\usepackage{wasysym}

\newcommand{\Hvec}{\mathbf{H}}
\newcommand{\Mvec}{\mathbf{M}}
\newcommand{\Dvec}{\boldsymbol{\Delta}}
\newcommand{\avec}{\mathbf{a}}
\newcommand{\Svec}{\mathbf{S}}
\newcommand{\Rvec}{\mathbf{R}}
\newcommand{\Pvec}{\mathbf{P}}
\newcommand{\Wvec}{\mathbf{W}}
\newcommand{\Jvec}{\mathbf{J}}
\newcommand{\Lvec}{\mathbf{L}}
\newcommand{\uvec}{\mathbf{u}}
\newcommand{\vvec}{\mathbf{v}}
\newcommand{\wvec}{\mathbf{w}}
\newcommand{\Ovec}{\mathbf{0}}
\newcommand{\muvec}{\boldsymbol{\mu}}
\newcommand{\etvec}{\boldsymbol{\eta}}
\newcommand{\svec}{\boldsymbol{\sigma}}
\newcommand{\svecvec}{\vec{\boldsymbol{\sigma}}}

\newcommand{\qvec}{\mathbf{q}}

\newcommand{\rvec}{\mathbf{r}}
\newcommand{\ham}{\mathcal{H}}
\newcommand{\NN}{\mathcal{N}}
\newcommand{\OO}{\mathcal{O}}
\newcommand{\adag}{a^\dagger}
\newcommand{\bdag}{b^\dagger}
\newcommand{\xx}{{\tilde{x}}}
\newcommand{\yy}{{\tilde{y}}}
\newcommand{\zz}{{\tilde{z}}}
\newcommand{\nn}{{\mathbf{\hat{n}}}}

\newcommand{\tr}{\mathrm{Tr \,}}
\newcommand{\half}{\frac{1}{2}}
\newcommand{\eff}{\mathrm{eff}}
\newcommand{\cl}{\mathrm{cl}}
\newcommand{\harm}{\mathrm{harm}}
\newcommand{\even}{\mathrm{even}}
\newcommand{\odd}{\mathrm{odd}}
\newcommand{\rref}{\mathrm{(ref)}}
\newcommand{\comb}[2]{\left( \begin{array}{c} #1 \\ #2 \end{array} \right)}
\newcommand{\tcomb}[3]{\left( \begin{array}{c} #1 \\ #2\, #3\end{array} \right)}
\newcommand{\ft}{\tilde{f}}
\newcommand{\gt}{\tilde{g}}
\newcommand{\llangle}{\langle \langle}
\newcommand{\rrangle}{\rangle \rangle}
\newcommand{\tdeta}{\tilde{\eta}}
\newcommand{\EE}{\mathcal{E}}
\newcommand{\BB}{\mathcal{B}}
\newcommand{\JJ}{\mathcal{J}}

\begin{document}

\title{Effective Hamiltonian for the Pyrochlore antiferromagnet: semiclassical 
derivation and degeneracy}
\author{U.~Hizi}
\email{uh22@cornell.edu}
\author{C.~L.~Henley}
\affiliation{Laboratory of Atomic and Solid State Physics,
Cornell University, Ithaca, NY, 14853-2501}

\begin{abstract} 
In the classical pyrochlore lattice Heisenberg antiferromagnet,
there is a macroscopic continuous ground state degeneracy.
We study semiclassical limit of large spin length $S$, keeping only the 
lowest order (in $1/S$) correction to the classical Hamiltonian.
We perform a detailed analysis of the spin-wave modes, and using a real-space
loop expansion, we produce an effective Hamiltonian, in which the degrees of 
freedom are Ising variables representing fluxes through loops in the lattice.
We find a family of degenerate collinear ground states, related by  gauge-like
$Z_2$ transformations and provide bounds for the order of the degeneracy. 
We further show that the theory can readily be applied to determine the ground
states of the Heisenberg Hamiltonian on related lattices,
and to field-induced collinear magnetization plateau states.
\end{abstract}


\maketitle

\section{\label{sec:intro}Introduction}

In recent years, there have been many theoretical studies of
geometrically frustrated systems
\cite{ramirez_review,diep}.
These are systems in which not all of the spin bonds can be satisfied 
simultaneously, due to the connectivity of the lattice.
The frustration can lead to unconventional magnetic ordering, 
or to a complete absence of long-range order. 

Of particular interest are nearest-neighbor antiferromagnets
on \emph{bisimplex} \cite{clh_heff}
lattices composed of corner sharing \emph{simplexes}\cite{moessner_01,clh_heff}.
Examples include the two-dimensional kagom\'e and the three-dimensional garnet
lattices, each composed of triangular simplexes, and lattices composed of
corner sharing ``tetrahedra'': the two-dimensional checkerboard, 
the layered SCGO~\cite{ramirez_review},
and the three-dimensional pyrochlore.

\begin{figure}[!h] 
\resizebox{7cm}{!}{\includegraphics{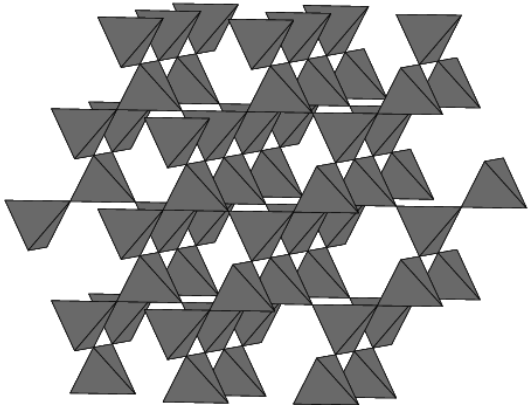}}
\caption{\footnotesize  The pyrochlore lattice \label{fig:lattice}}
\end{figure}

The pyrochlore lattice, in which the centers of the tetrahedra form a diamond 
lattice\cite{footnote_1}
(see Fig.~\ref{fig:lattice}), is of interest because it 
is realized in many experimental systems, in both $\mathrm{A_2 B_2 O_7}$ oxides
and in B sites of $\mathrm{A B_2 O_4}$ spinels~\cite{greedan_review},
and because, by the analysis of Ref.~\onlinecite{moessner}, bisimplex
lattices composed of tetrahedra are
less susceptible to ordering than lattices with triangle simplexes.

We consider the semiclassical limit of the 
pyrochlore nearest neighbor Heisenberg Hamiltonian
\begin{equation} \label{eq:ham}
\ham = \sum_{i j} J_{ij} \Svec_i \cdot \Svec_j \,,
\end{equation}
where $J_{ij}\!=\!1$ for nearest neighbors $\langle i j \rangle$.
and $\Svec_i$ is a spin of length $S\!\gg\! 1$ at site $i$.
The Hamiltonian can be recast in the form
\begin{equation}
\ham= \sum_\alpha | \Lvec_\alpha |^2 + \mathrm{const.} \,,
\end{equation}
where $\Lvec_\alpha \!\equiv\!\sum_{i \in \alpha} \Svec_i$, is an operator that
lives on the diamond lattice sites 
$\alpha$ (we reserve Greek indices for tetrahedra -
diamond lattice sites - 
and roman indices for pyrochlore sites).
The result is that, classically, all states satisfying
\begin{equation} \label{eq:cl_const}
\Lvec_\alpha =0\,.
\end{equation}
for all tetrahedra $\alpha$, are degenerate classical ground states.
Eq.~(\ref{eq:cl_const}) has three ($L_\alpha^x$,$L_\alpha^y$,$L_\alpha^z$)
constraints per tetrahedron, i.e. $3/2$ constraints per site.
Since there are two degrees of freedom (spherical angles $\theta$,$\phi$) per
site, there remain an extensive number of degrees of freedom, and a 
degeneracy that is exponential in an extensive quantity~\cite{moessner}.
This type of massive continuous degeneracy has been shown in the 
kagom\'e lattice antiferromagnet to be lifted
due to thermal\cite{chalker,huse} or (anharmonic) 
quantum~\cite{chubukov+chan}
fluctuations, a phenomenon known as
\emph{order by disorder}~\cite{villain,clh_obd}.
However, in the classical case on the pyrochlore model, thermal fluctuations
are not believed to facilitate ordering~\cite{reimers,moessner}. 

In the $S\!\gg\!1$ quantum model, one of us has 
demonstrated\cite{clh_harmonic} that the classical degeneracy is not fully
lifted by the lowest order (in $1/S$) non-interacting spin-wave theory, 
assuming a collinear spin arrangement.
In this paper, we recover this result using the more rigorous
Holstein-Primakoff transformation, and provide a detailed study of various 
aspects in the linear spin-wave theory. 

Recently, there have been various works designed 
to search for a ground state of the pyrochlore in the large-$S$ limit.
These include work on two-dimensional analogues of the
pyrochlore~\cite{capped_kagome,tcher_check}
(see also Sec.~\ref{sec:other_lattices}),
on ground state selection due to lattice distortion and spin-orbit
coupling in Vanadium spinels~\cite{vanadium},
and on the large-$N$ mean field theory~\cite{uh_LN}.
Another body of work is on the closely related problem of the ordering
in the pyrochlore in the 
presence of a magnetic field, which induces a collinear
pattern~\cite{penc+bergman} (see also Sec.~\ref{sec:field}).

The rest of this paper is organized as follows:
In Sec.~\ref{sec:model} we derive the large-$S$ expansion of the
Hamiltonian~(\ref{eq:ham})
\begin{equation} \label{eq:ham_exp}
\ham = \ham_\cl + \ham_{\harm} + \ham_{\mathrm{quart}} + \ldots \,,
\end{equation}
where $\ham_\cl$ is the classical Hamiltonian , of order $S^2$,
$\ham_\harm$ is the harmonic, non-interacting spin-wave, Hamiltonian
(of order $S$) and $\ham_\mathrm{quart}$ is the next order ($\OO(1)$)
interaction term~\cite{kvale}.
We diagonalize the Harmonic part of the expansion,
assuming fluctuation around a \emph{collinear} classical ground state,
where each state can be parameterized by an Ising variable $\eta_i=\pm 1$
on each lattice site, such that $\Svec_i=\eta_i \hat{z}$.

In Sec.~\ref{sec:modes} we study the properties of the various spin-wave modes:
\emph{zero modes} that do not contribute to the zero-point energy, 
\emph{non-zero modes} that can be expressed entirely in terms of the diamond
lattice sites,
and \emph{divergent zero modes} that carry divergent fluctuations.
We use this formalism to 
demonstrate a key result of Ref.~\onlinecite{clh_harmonic}:
that collinear classical ground states related by Ising \emph{gauge-like}
transformations are exactly degenerate.

In Sec.~\ref{sec:zeropoint} we consider the zero-point energy of the Harmonic 
fluctuations and derive an \emph{effective Hamiltonian}~\cite{clh_heff}, where
we parameterize the energy only in terms of \emph{flux} variables through each
\emph{loop} in the diamond lattice
(where the pyrochlore spins sit bond centers).
\begin{equation} \label{eq:heff1}
E_\harm^\eff = E_0 + K_{6} \Phi_{6} + K_{8} \Phi_{8} 
+\ldots \,,
\end{equation}
where $K_n$ are numerical coefficients that we evaluate using a real-space
\emph{loop expansion},
and $\Phi_{n}$ are sums of products of the Ising variables $\eta_i$ around loops
of length $n$. 
We numerically evaluate the zero-point energy of a large number of collinear
classical ground states and find that the effective Hamiltonian does a good
job of evaluating the energy with just a few terms.
We find a family of ground states and find an upper limit to the entropy,
using a correspondence between gauge-like transformations and divergent zero
modes.

In Sec.~\ref{sec:noncoll} we demonstrate that the collinear states have lower
energy than non-collinear states obtained by rotating loops of spins
out of collinearity, thereby showing that our assumption of collinearity was
justified.
In Sec.~\ref{sec:other_models} we apply the loop expansion to some closely 
related models including the case of non-zero magnetization plateaus and
other lattices. We find that the effective Hamiltonian approach is useful in
predicting the ground states in many cases.
In Sec.~\ref{sec:concl} we present our conclusions.

\section{Semiclassical spin-waves}\label{sec:model}

In this section, we consider the effect of quantum fluctuations on the
pyrochlore Heisenberg model, in the semiclassical, $S \gg 1$ limit.
In~\ref{sec:large_S} we perform a Holstein Primakoff transformation to
expand the Hamiltonian in powers of $1/S$.
In~\ref{sec:collinear}, we focus on collinear classical ground states,
Next, in Sec.~\ref{sec:diag}, we diagonalize the collinear harmonic Hamiltonian
to find the spin-wave modes. 

\subsection{Large-$S$ expansion} \label{sec:large_S}
We start from a given ordered classical state, where the spin directions are
parameterized by angles $(\theta_i, \phi_i)$, such that the classical spin 
direction is
\begin{equation}
\nn_i =
(\cos \theta_i, \sin \theta_i \cos \phi_i, \sin \theta_i \sin \phi_i)\,.
\end{equation}
We expand around this state, in
powers of $1/S$, to account for  quantum fluctuations,
implicitly assuming here that the quantum fluctuations are small and do
not destroy the local collinear order.
Upon rotation to local axes $(\xx,\yy,\zz)$, such that the classical spins
are in the $\zz$ direction (parallel to $\nn_i$), we apply
the usual Holstein-Primakoff transformation.
Note that the choice of directions $\xx$ and $\yy$ is arbitrary, and in the
following we shall take $\yy$ to be perpendicular to the $z$ axis.
One defines boson operators $a_i$, $\adag_i$ such that
\begin{eqnarray}
S^\zz_i &=& S- \adag_i a_i \,, \nonumber \\
S^{+}_i &\equiv&  S^\xx + i S^\yy =
\sqrt{2S-\adag_i a_i} \,\, a_i \approx \sqrt{2S} a_i
\,, \nonumber \\
S^{-}_i &\equiv&  S^\xx - i S^\yy =
\adag_i \sqrt{2S-\adag_i a_i}
\approx \sqrt{2S} \adag_i \,.
\label{eq:holstein}
\end{eqnarray}
These operators satisfy the canonical bosonic commutation relations
\begin{equation}
[a_i,\adag_j] =\delta_{ij} \,, \qquad
[a_i,a_j] =0 \,, \qquad
[\adag_i,\adag_j] =0 \,.
\end{equation}
We obtain the Hamiltonian~(\ref{eq:ham_exp}), 
where the leading term is the classical Hamiltonian
\begin{equation} \label{eq:hcl}
\ham_\cl= S^2 \sum_{i j} J_{ij}
\nn_i \cdot \nn_j \,,
\end{equation}
which  is equivalent of Eq.~(\ref{eq:ham}), and
whose degenerate ground states satisfy
\begin{equation}
\label{eq:classical_gs}
\sum_{i \in \alpha} \nn_i = \mathbf{0} \,.
\end{equation}
for all tetrahedra $\alpha$.
Due to the large classical degeneracy, we must go on to the leading order,
harmonic, quantum correction, in order to search for a ground state, 
while assuming that Eq.~(\ref{eq:classical_gs}) is satisfied, i.e., that we
are expanding around a classical ground state.
The linear spin-wave energy $E_\harm$ was calculated
by one of us~\cite{clh_harmonic}, in previous work, 
using classical equations of motion.
The results of that work were that the classical degeneracy is not fully
lifted by quantum fluctuations, to harmonic order, and that the remaining
degeneracy is associated with a \emph{gauge-like} symmetry.
In this paper we justify these results using the more rigorous
Holstein-Primakoff approach, which allows us to gain a better analytic
understanding of the degeneracy, as well as perform numerical diagonalization.
Furthermore, the Holstein-Primakoff transformation allows for a controlled
expansion in powers of $1/S$, including anharmonic order. 
In following work, we intend to show that the interacting spin-wave theory
breaks the harmonic degeneracy~\cite{uh_quartic}.

We find it convenient to change variables to \emph{spin deviation operators}
\begin{equation}
\svec^\xx = \sqrt{\frac{S}{2}} (\avec+\avec^\dagger) \,, \qquad
\svec^\yy= -i \sqrt{\frac{S}{2}} (\avec-\avec^\dagger) \,,
\end{equation}
where $\avec^\dagger=(\adag_1,\adag_2,\ldots,\adag_{N_s})$
is a vector of operators of length $N_s$ (the number of sites).
We shall, from now on, reserve boldface notation for such vectors and
matrices.
These operators satisfy the commutation relations
\begin{equation}
[\svec^\xx,\svec^\yy]=i S \openone \,, \qquad
[\svec^\xx,\svec^\xx]=[\svec^\yy,\svec^\yy]=0 \,.
\end{equation}
The harmonic Hamiltonian can now be written in matrix notation
\begin{equation}
\ham_\harm=
\left((\svec^\xx)^\dagger ,(\svec^\yy)^\dagger \right)
\left( \begin{array}{cc}
\Rvec^\xx & \Pvec \\
\Pvec^T & \Rvec^\yy \end{array} \right)
\left( \begin{array}{c}
\svec^\xx \\ \svec^\yy \end{array} \right)-
S N_s \,,
\label{eq:harm_mat}
\end{equation}
where the block matrixes, with respect to lattice site index are
\begin{eqnarray}
P_{ij} &=& \frac{J_{ij}}{2} \cos \theta_i \sin \phi_{ij} \,, \nonumber \\
R^\xx_{ij} &=& \delta_{ij}+\frac{J_{ij}}{2} \left(\sin \theta_i \sin \theta_j +
\cos \theta_i \cos \theta_j \cos \phi_{ij}\right) \,, \nonumber \\
R^\yy_{ij} &=& \delta_{ij}+\frac{J_{ij}}{2} \cos \phi_{ij} \,,
\end{eqnarray}
where we defined $\phi_{ij}\!\equiv\! \phi_i-\phi_j$.
These matrices depend on our arbitrary choice for the local transverse directions
$\xx$ and $\yy$, and can therefore not be expressed solely in terms of the spin
direction $\nn_i$.
Note however, that in the case of \emph{coplanar} spins,
one can take, with no loss of generality, $\phi_i\!=\!0$ for all sites, i.e.
spins in the $(x,z)$ plane, and find that $P_{ij}\!=\!0$ and that the matrix
in Eq.~(\ref{eq:harm_mat}) is block diagonal.

\subsection{Harmonic Hamiltonian for Collinear states} \label{sec:collinear}

The preceding derivation (Eqs.~(\ref{eq:ham}-\ref{eq:harm_mat})) is valid for
any lattice composed of corner sharing simplexes. 
One can argue on general grounds~\cite{shender,clh_obd,clh_heff,larson+zhang}
that the spin-wave energy has local minima for classically collinear states.
In lattices that are composed of corner-sharing triangles, such as the kagom\'e,
the constraint (Eqs.~(\ref{eq:cl_const},\ref{eq:classical_gs})) is incompatible
with collinearity.
However, in the case of the pyrochlore lattice,
or any other lattice composed of corner sharing tetrahedra,
there is an abundance of collinear classical ground states.
We will henceforth assume a collinear classical ground state, i.e.
that each site is labeled by an Ising variable
$\eta_i\! =\! \pm 1$, such that $\Svec_i = \eta_i \hat{z}$. 
We will return to the more general, non-collinear case in Sec.~\ref{sec:noncoll}
to justify this assumption a posteriori.

In the collinear case ($\phi_i\!=\!0$, $\theta_i\!\in\! \{0,\pi\}$), 
our definitions of the local axes for site $i$, are such that
\begin{equation}
\xx = \eta_i x \,, \qquad \yy=y \,, \qquad \zz=\eta_i z \,.
\end{equation}
We may restore the $x$-$y$ symmetry of the problem, by 
transforming the spin deviation operators back to the regular axes by the 
reflection operation
\begin{equation} \label{eq:sdev}
\svec^x \equiv \etvec \svec^\xx \,,\qquad \svec^y \equiv \svec^\yy \,, \qquad
\svecvec \equiv \left( \begin{array}{c} \svec^x \\ \svec^y
\end{array} \right) \,,
\end{equation}
where
\begin{equation}
\etvec \equiv \left(
\begin{array}{cccccc}
\eta_1 & 0 & 0 & \cdots & 0 \\
0 & \eta_2 & 0 & \cdots & 0 \\
0 & 0 & \eta_3 & \cdots & 0 \\
\vdots & \vdots & \vdots &\ddots & \vdots \\
0 & 0 & 0 & \cdots & \eta_N
\end{array}
\right) \,.
\end{equation}
The transformation~(\ref{eq:sdev}) amounts to a reflection with respect to the
$y$-$z$ plane and makes the diagonal blocks in Eq.~(\ref{eq:harm_mat}) equal to
each other:
\begin{equation}
\ham_\harm= \svecvec^\dagger \left(
\begin{array}{cc}
\Hvec & 0 \\ 0 & \Hvec \end{array} \right) \svecvec - S \, \tr \Hvec\,,
\label{eq:coll_harm_mat}
\end{equation}
where
\begin{equation}
\label{eq:heis_mat}
\Hvec = \openone + \frac{\Jvec}{2} =  \half \Wvec^\dagger \Wvec 
\,,
\end{equation}
$\Jvec$ is the matrix whose elements are $J_{ij}$ and
$\Wvec_{\alpha i}$ is an $N_s/2 \!\times\! N_s$ matrix that takes the
value $1$ if  $i\in \alpha$, and $0$ otherwise.

The transformation of Eq.~(\ref{eq:sdev}) was chosen to explicitly
show the symmetry between the $x$ and $y$ axes in the collinear case.
It also causes the 
Hamiltonian~(\ref{eq:coll_harm_mat}) to appear independent of $\{\eta_i\}$.
However, the particular collinear configuration does enter the calculation via
the commutation relations
\begin{equation} \label{eq:sig_comm}
[\svec^x,\svec^y] = i S \etvec
\end{equation}
Therefore, the equations of motion that govern the spin-waves \emph{do}
depend on the classical ground state configuration.

\subsection{Diagonalization of harmonic Hamiltonian} \label{sec:diag}

Next we would like to Bogoliubov-diagonalize the Hamiltonian of
Eq.~(\ref{eq:coll_harm_mat}), so that we can study its eigenmodes and
zero-point energy.
As a motivation, we write the equations of motion
\begin{equation} \label{eq:motion}
\frac{d \svecvec}{d t} = -i[\svecvec,\ham_\harm]=
2S
\left(
\begin{array}{cc}
0 &  \etvec \Hvec \\
-\etvec \Hvec & 0
\end{array}
\right) \svecvec \,.
\end{equation}
These are the quantum equivalent of the classical equations of motion derived
in Ref.~\cite{clh_harmonic}.
Upon Fourier transforming with respect to time, and squaring the matrix,
we obtain
\begin{equation} \label{eq:motion2}
\left(\frac{\omega}{2S}\right)^2 \svecvec  =
\left(
\begin{array}{cc}
(\etvec \Hvec)^2 & 0 \\
0 & (\etvec \Hvec)^2
\end{array}
\right) \svecvec \,.
\end{equation}
The spin-wave modes are therefore eigenvectors of the
matrix $(\etvec \Hvec)^2$, with eigenvalues $\lambda_m^2\!=\!(\omega_m/2S)^2$.
If the non-hermitian matrix $\etvec \Hvec$ is diagonalizable then its
eigenvectors $\{\vvec_m\}$ are the spin-wave modes with eigenvalues
$\{\pm|\lambda_m|\}$. Here $\vvec_m$ is a vector of length $N_s$, 
whose indices are site numbers.
Note that although we refer to the eigenvectors of $\eta \Hvec$ as the spin-wave
modes, strictly speaking, the quantum mechanical modes are pairs of conjugate 
operators $\svec^x\!=\!\vvec_m$,
$\svec^y\!=\!i \mathrm{sgn}(\lambda_m) \vvec_m$.

Since $\{ \Hvec^{1/2} \vvec_m \}$ are eigenvectors of the Hermitian matrix
$\Hvec^{1/2} \etvec \Hvec^{1/2}$, the complete basis of
eigenvectors $\{\vvec_m\}$ can be ``orthogonalized'' by
\begin{equation}
(\vvec_m, \etvec \vvec_n) \equiv \vvec_m^\dagger \etvec\vvec_n = 
\frac{1}{\lambda_m} \vvec_m^\dagger \Hvec \vvec_n =
c_m \delta_{m,n} 
\,. \label{eq:orth}
\end{equation}
We shall henceforth refer to the operation $(\vvec_m, \etvec \vvec_n)$
as the \emph{inner product} of modes $\vvec_m$ and $\vvec_n$.
Note that $c_m\!\equiv\!(\vvec_m,\etvec \vvec_m)$ is not really a norm,
since it can be zero or, if $\lambda_m\!<\!0$, negative.
The Bogoliubov diagonalization involves transforming to boson operators
\begin{equation}
b_m = \frac{1}{\sqrt{2 S |c_m|}}
\left( (\etvec \vvec_m)^\dagger \svec^x 
+ i \,\mathrm{sgn}(c_m) (\etvec \vvec_m)^\dagger \svec^y 
\right) \,, 
\end{equation}
with canonical boson commutation relations, to obtain
\begin{equation}
\ham_\harm =  \sum_m \omega_m \left(\bdag_m b_m + \frac{1}{2} \right) - S N_s
\,,
\label{eq:harm}
\end{equation}
with the zero-point energy
\begin{equation}
\label{eq:zeropoint}
E_\harm=\half \sum_m \omega_m =S \sum_m |\lambda_m| -SN_s\,.
\end{equation}
The fluctuations of $\svec^{x,y}$ are now easy to calculate from the 
boson modes
\begin{eqnarray}
\langle \svec^x (\svec^x)^\dagger \rangle =
\langle \svec^y (\svec^y)^\dagger \rangle 
&=& \sum_m 
\frac{S}{2|c_m|} 
\vvec_m \vvec_m^\dagger \,, \nonumber \\
\langle \svec^x (\svec^y)^\dagger +  \svec^y (\svec^x)^\dagger \rangle
&=& \Ovec \,.
\label{eq:fluct}
\end{eqnarray}
Note that this is a matrix equation, where
$\svec^{x/y} (\svec^{x/y})^\dagger$ are $N_s \! \times \! N_s$ matrices.
%

\section{Spin-wave modes} \label{sec:modes}

We now examine the spin-wave modes that we found in Sec.~\ref{sec:diag} to 
study their properties.
In~\ref{sec:zero} we classify them as zero modes,
that do not contribute to the zero-point energy,
and non-zero modes, that can all be expressed in terms of the diamond lattice
formed from the centers of tetrahedra.
Within the zero modes, we find, in~\ref{sec:div},
that a number of modes (proportional to $N_s^{1/3}$)
have divergent fluctuations.
In Sec.~\ref{sec:band} we discuss the energy band  structure
and identify special singular lines in the Brillouin zone.
In~\ref{sec:gauge}, we show that the energy of any non-zero
mode is invariant under an group of $Z_2$ gauge-like transformations.

\subsection{Zero modes and Non-zero modes} \label{sec:zero}

\begin{figure}
(a) \resizebox{!}{3cm}{\includegraphics{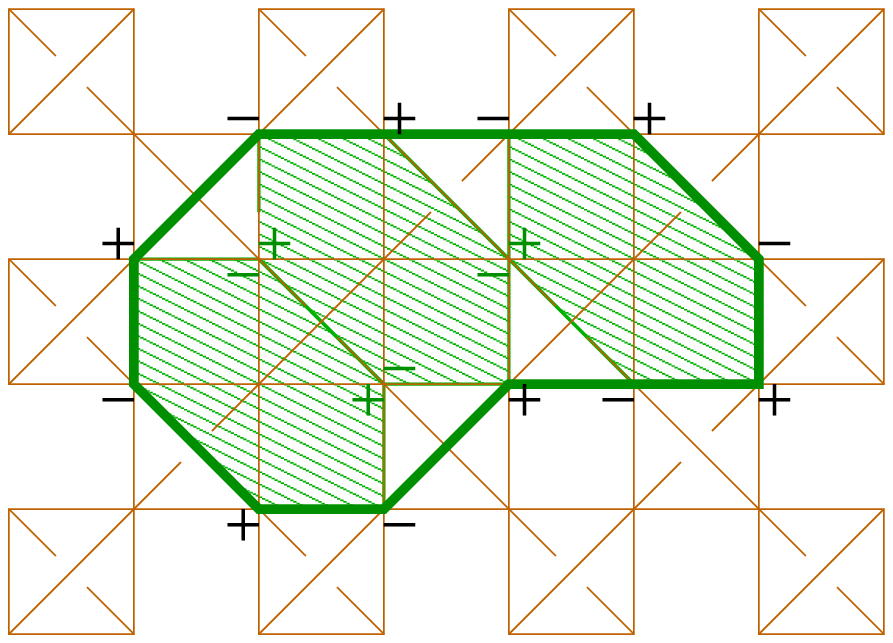}}
(b) \resizebox{!}{3cm}{\includegraphics{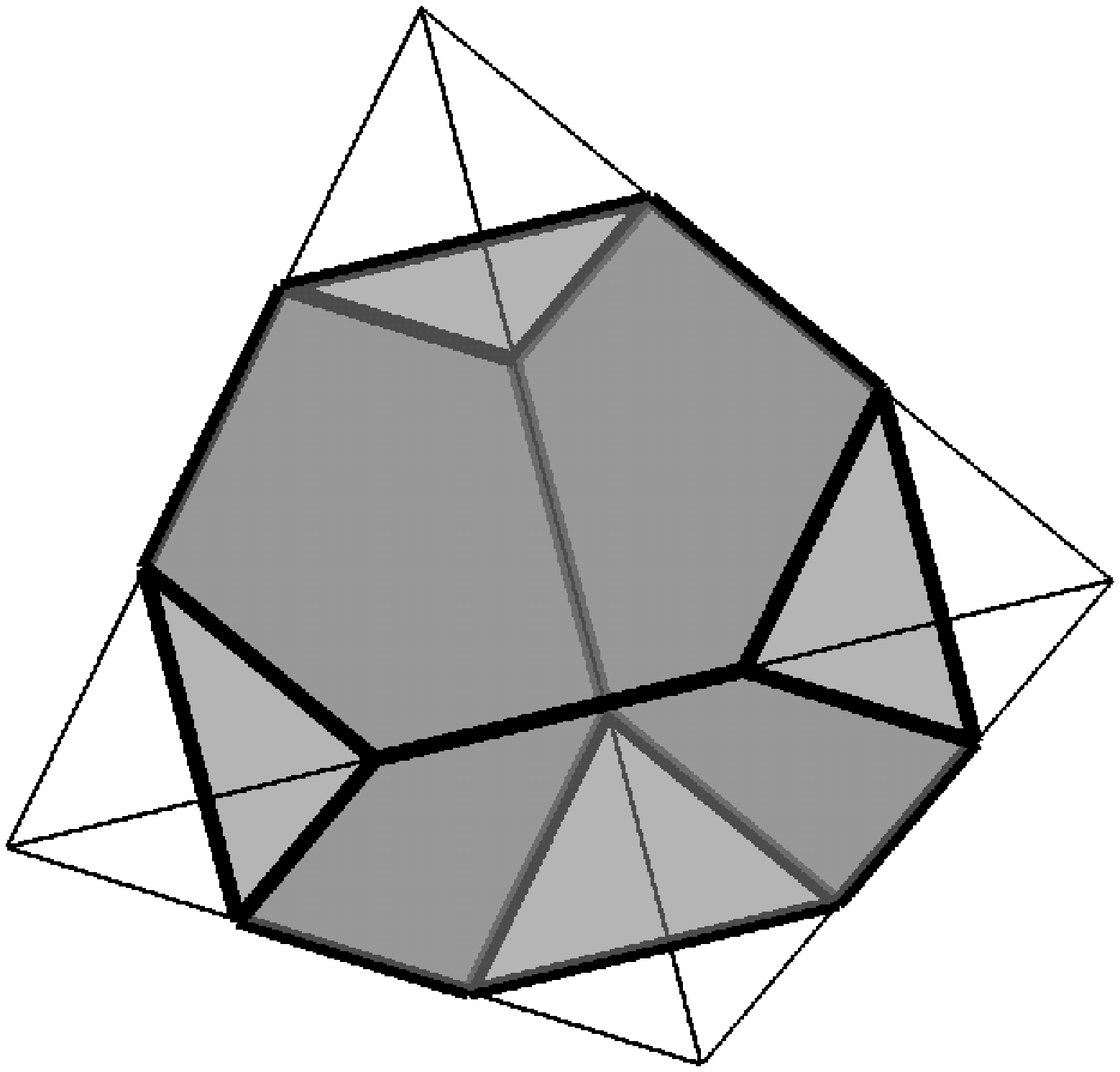}}
\caption{ \label{fig:zero_modes} \footnotesize (Color Online)
(a) Example of a zero mode (in (001) projection) 
alternating along a loop. 
Any such loop can be expressed as a linear combination of $N_s$ hexagon 
modes. Furthermore, there is a linear dependence between
the hexagons in each super-tetrahedron, as depicted in (b).
The additional constraints reduce the number of independent modes to $N_s/2$.
}
\end{figure}

Many of the eigenmodes of the Hamiltonian~(\ref{eq:heis_mat}),
are \emph{zero modes}, i.e. modes that are associated a frequency
$\omega_m\!=\!0$.
These are the modes $\vvec_z$ that satisfy
\begin{equation}
(\Wvec \vvec_z)_\alpha=
\sum_{i\in \alpha} \vvec_z(i) =0 \qquad \mbox{for all tetrahedra } \alpha \,.
\end{equation}
Since there are $N_s/2$ tetrahedra, and thus $N_s/2$ constraints,
we can expect as many as 
half of the spin-wave frequencies $\{\omega_m\}$ to be generically zero.
In order to prove that indeed half of the eigenmodes of $\Hvec$ are zero modes
we first note that the subspace of zero modes is
spanned by modes that alternate along \emph{loops},
as in Fig.~\ref{fig:zero_modes}, i.e., for a loop $\mathcal{L}$,
up to a normalization factor
\begin{equation}
v_\mathcal{L}(i) \propto \left\{ \begin{array}{ll}
(-1)^{n_i} &  i \mbox{ site number } n_i \mbox{ in loop } \mathcal{L} \\
0 & i \notin \mathcal{L}
\end{array} \right. \,.
\end{equation}
In order for these loop modes to indeed be zero modes, two consecutive bonds
in the loop cannot be from the same tetrahedron,
i.e. the sites along each loop are centers of bonds in a diamond lattice loop. 
From now on, we will use the term \emph{loop} only for lines in the lattice 
satisfying this constraint.
The loop zero modes can in turn be written as a linear
combination of hexagon modes only, of which there are $N_s$
(see Fig.~\ref{fig:zero_modes}a).
Hexagons are the shortest loops in this lattice.
Since there is a linear dependence between the four hexagons in a big 
super-tetrahedron (see Fig.~\ref{fig:zero_modes}b), then only $N_s/2$ of the 
hexagon modes are linearly independent~\cite{moessner,zhit}.
Therefore a basis of zero modes  of $\Hvec$ would consist of half of the 
hexagon modes, and the matrix
$(\etvec \Hvec)^2$ has \emph{at least} $N_s/2$ zero modes.
Note that since these modes are zero modes of the matrix $\Hvec$, 
they are independent of the particular spin arrangement.
We refer to these $N_s/2$ modes as the \emph{generic zero modes}, to 
distinguish them from other zero modes of $(\etvec \Hvec)^2$ that are not 
zero modes of $\Hvec$, that we shall discuss in the next section.
\emph{All collinear ground states have the same generic zero modes}.

Zero modes mean that the
quadratic correction to the classical energy due to small deviations 
from collinear order vanishes.
It is interesting to note, that the zero modes associated with loops
that have alternating classical spins $\eta_i$,
are modes associated with the transformations under which $\ham_\cl$ is
invariant.
In fact, deviations that rotate spins along these loops by any angle $\theta$
take a collinear state to a non-collinear classical ground state, and 
when $\theta\!=\!180^\circ$, the rotation takes one collinear classical ground
state to another.
These modes are completely analogous to the so-called \emph{weathervane modes}
in the kagom\'e lattice~\cite{chandra}.
We will use the properties of these rotations in Sec.~\ref{sec:numerical}
to generate a large number of collinear classical ground state, and
in Sec.~\ref{sec:noncoll} to study the dependence of the zero-point energy 
on deviations out of collinearity.

Thus, the abundance of spin-wave zero modes is a reflection of the
macroscopic continuous classical degeneracy.
On the experimental side,
the properties of the spinel material $\mathrm{ZnCr_2O_4}$, at 
temperatures just higher than the phase
transition into an ordered state, have been shown to be
dominated by so-called \emph{local soft modes}, which are spin-wave modes that
have zero or infinitesimal frequency~\cite{shlee_nature}.

Whereas the zero modes are associated with the large classical degeneracy,
they do not contribute to the quantum zero-point energy~(\ref{eq:zeropoint}).
We now consider harmonic modes that have non-zero frequency.
For any such mode $\vvec_{nz}$, the diamond lattice vector
$\uvec= \Wvec \vvec_{nz}$ is an eigenvector of
$(\Wvec \etvec \Wvec^\dagger )^2$ with the same eigenvalue
\begin{equation} \label{eq:eig_tet}
(\Wvec \etvec \Wvec^\dagger )^2 \uvec =
\Wvec (\etvec \Wvec^\dagger \Wvec)^2 \vvec_{nz}=
\lambda^2 \uvec
\end{equation}
In this fashion, we can get rid of $N_s/2$ generic zero modes by projecting
to a space that lives on the diamond lattice sites only, and is orthogonal to
all generic zero modes.
As already noted, since the generic zero modes are the same for all states,
we can
consistently disregard them and limit ourselves solely to diamond lattice
eigenmodes of Eq.~(\ref{eq:eig_tet}).
Since the matrix $\Wvec \etvec \Wvec^\dagger$ is symmetric, it
is always diagonalizable and its eigenvectors are orthogonal in the standard
sense.
\begin{equation} \label{eq:u_orth}
\uvec_m \cdot \uvec_n \propto \delta_{mn}\,.
\end{equation}
We refer to these remaining $N_s/2$ spin-wave modes,
that can be viewed as diamond lattice modes, as the \emph{ordinary modes}.
Although the ordinary modes do not generically have zero frequency, we
may find that for a given classical ground state, some of them are zero modes.
We will find, in the following section, that these are modes that have divergent
fluctuations.

\subsection{Divergent modes} \label{sec:div}

\begin{figure}[!h] 
(a) \resizebox{4.3cm}{!}{\includegraphics{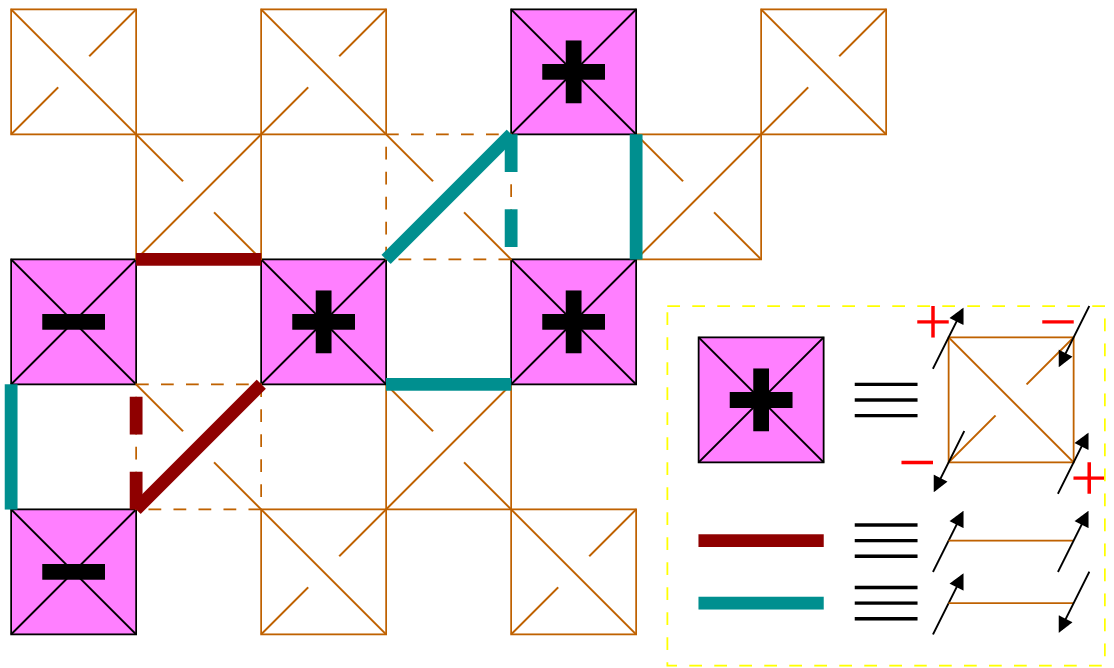}}
(b) \resizebox{3.2cm}{!}{\includegraphics{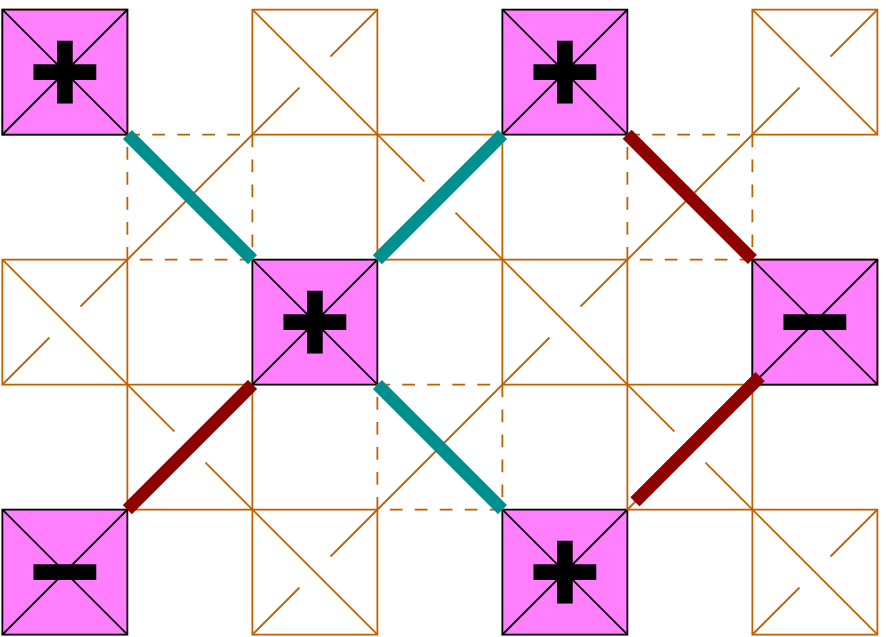}}
\caption{\label{fig:div}
\footnotesize (Color online) (a) Example of a portion of a (001) slice out of
an even divergent mode, in real space
(dashed lines connect to adjacent slices).
Light (dark) colored bonds represent satisfied (unsatisfied) antiferromagnetic
bonds, and an even
tetrahedron $\alpha$ is marked ``$\pm$'' for $u_d(\alpha)\! =\! \pm 1$.
Each participating tetrahedron 
must be have four others so that Eq.~(\ref{eq:div_const}) is satisfied
for all (even and odd) tetrahedra.
Thus two even tetrahedra $\alpha$, $\gamma$
have $u_d(\alpha)\!=\!u_d(\gamma)$ if they are 
connected by a satisfied bond, and $u_d(\alpha)\!=\! - u_d(\gamma)$ otherwise.
(b) Example of a planar $(xy)$ divergent mode, i.e. a mode that is bounded along
the $z$ axis (in this case, to one slice).
Divergent modes can be bounded at most, along one of the major axes.}
\end{figure}

As is apparent in Eq.~(\ref{eq:fluct}),
divergent harmonic fluctuations occur whenever a certain mode $\vvec_d$
satisfies
\begin{equation}
c_d\equiv (\vvec_d, \etvec \vvec_d) = 0\,.
\end{equation}
This can be shown to occur if and only if $\etvec \Hvec$ is not diagonalizable,
i.e., when the Jordan form of the matrix has a block of order 2.
\begin{eqnarray}
\etvec \Hvec \vvec_d &=& 0 \nonumber \,, \\
\etvec \Hvec \wvec_d &=& \vvec_d \,.
\end{eqnarray}
We call the mode $\wvec_d$, which is a zero mode of $(\etvec \Hvec)^2$,
but not an eigenmode of $\etvec \Hvec$, a \emph{divergent zero mode}.
As we noted before, a more accurate description would be to refer to the
pair of conjugate operators  $\svec^x \!\sim\!\vvec_d$,
$\svec^y \!\sim\! \wvec_d$ as the quantum mechanical divergent mode.
From this, using Eq.~(\ref{eq:heis_mat}), we find that
\emph{any} divergent zero mode is related to a 
zero mode of Eq.~(\ref{eq:eig_tet}), which is defined on the diamond lattice
\begin{equation}
\uvec_d \equiv \Wvec \wvec_d \,.
\end{equation}
We call $\uvec_d$ a \emph{diamond lattice zero mode}.
Diamond lattice zero modes can be constructed so that they are restricted to
only one of the two (FCC) sublattices.
For example, one can construct an even divergent mode $\uvec_d$, i.e.
\begin{equation} \label{eq:div_ev}
u_d(\alpha)\in \{0, \pm 1\} \mbox{ for even }\alpha \,, 
u_d(\alpha) = 0 \mbox{ for odd }\alpha\,,
\end{equation}
such that for each of the adjacent odd sites $\beta$ 
\begin{equation} \label{eq:div_const}
\sum_{\gamma: \langle \gamma \beta \rangle} u_d(\gamma) \eta_{i(\gamma,\beta)}
= 0 \,.
\end{equation}
Here $i(\gamma,\beta)$ is the pyrochlore site that belongs to both $\gamma$ and
$\beta$.
The mode is constructed in the following way: in addition to $\alpha$, one 
other, freely chosen, (even) nearest neighbor $\gamma$ of $\beta$ must be in 
the support of $\uvec_d$.
If the bond connecting $\alpha$ and $\gamma$ is satisfied (antiferromagnetic)
$u_d(\gamma)\!=\!u_d(\alpha)$. Otherwise, $u_d(\gamma) \! = \! -u_d(\alpha)$
(see Fig.~\ref{fig:div}).
We repeat this until the mode span the system size, in the percolation sense,
and all odd tetrahedra satisfy Eq.~(\ref{eq:div_const}).
We call this an \emph{even divergent mode}, 
and \emph{odd divergent modes} are constructed similarly from the odd
sublattice. 
Even and odd divergent modes are linearly independent from each
other, and
any divergent modes can be separated to a sum of an even divergent mode 
and an odd divergent mode, and since odd and even modes
are linearly independent from each other,
one can construct a real-space basis for divergent modes from the
single-sublattice divergent modes.

If we examine the rules of constructing real space divergent modes,
we find that their support (on the diamond lattice)
can either be unbounded in space, 
or bounded along \emph{only one} of the major axes, as in Fig.~\ref{fig:div}b.
This can be easily proved:
suppose that an even divergent mode is bounded in the $x$ direction.
There is an even diamond lattice site $\alpha$ that is in the support of 
this divergent mode, and its $x$ value takes the maximum possible
value $x_\alpha$
Note that in a slight abuse of notation, here ($x$,$y$,$z$) are spatial
coordinates, while in Sec.~\ref{sec:model} they referred to spin directions.
The site $\alpha$ has an odd neighbor
at $(x,y,z)\!=\!(x_\alpha\!+\!1/4,y_\alpha\!+\!1/4,z_\alpha\!-\!1/4)$ (where
the lattice constant of the underlying cubic lattice is taken to be $1$), 
that has two (even) neighbors at $x\!>\!x_\alpha$ (and thus cannot be in 
the support of the mode), and two neighbors at $x\!=\!x_\alpha$.
One of these is $\alpha$, and by the rules of the construction the other, 
at $(x_\alpha,y_\alpha\!+\!1/2,z_\alpha\!-\!1/2)$,
must also be in the support of the mode.
Continuing this reasoning, we will find that the site at coordinates
$(x_\alpha,y_\alpha\!+\!1,z_\alpha\!-\!1)$ is also in
the support of the divergent mode, and by induction the support of this mode
is unbounded in the $+y$ and
$-z$ directions.
Similarly, we can show that the support for this mode is unbounded in the $-y$
axis and the $+ z$ directions.

\subsection{Magnon energy band structure} \label{sec:band}

Our discussion so far has focused on spin waves in real space, and has therefore
been applicable to any (even non-periodic) classical collinear ground state.
However, in order to perform numerical calculation
(in Sec.~\ref{sec:numerical}), one must consider periodic spin configurations,
with a (possibly large) \emph{magnetic unit cell}.
We refer to the lattice composed of the centers of the magnetic unit cells as 
the \emph{magnetic lattice}.
We can Fourier transform the Hamiltonian~(\ref{eq:coll_harm_mat}), using
\begin{eqnarray}
\vec{\sigma}^l_{\rvec}
 &=& \frac{1}{\sqrt{N_M}} \sum_{\qvec} \vec{\sigma}^l_{\qvec}
 e^{-i \qvec \cdot (\rvec + \Dvec_l)}\,,\nonumber \\
 \vec{\sigma}^l_{\qvec} &=& \frac{1}{\sqrt{N_M}}
 \sum_{\rvec} \vec{\sigma}^l_{\rvec}
 e^{i \qvec \cdot (\rvec + \Dvec_l)}\,,
\end{eqnarray}
where $\rvec$ is a magnetic lattice vector, $l$ is a sublattice index, 
corresponding to a basis vector $\Dvec_l$, $N_M$ is the number of magnetic
lattice sites, and $\qvec$ is a Brillouin zone vector.
The elements of the transformed  Hamiltonian matrix are
\begin{equation}
H_{lm}(\qvec) = \delta_{lm} + \half \sum_{\xi_{lm}}
e^{i \qvec \cdot \xi_{lm}}\,,
\end{equation}
where the sum is over all nearest neighbor vectors $\xi_{lm}$ connecting
the sublattices $l$ and $m$.
Upon diagonalization, we obtain that the number of bands in the Brillouin zone
is equal to the number of sites in the magnetic unit cell,
i.e. the number of sublattices.
The bands can be classified as follows:
half of the energy bands belong to generic zero modes,
which have vanishing energy throughout the Brillouin zone.
These modes are the zero modes of $\ham$ and are identical for all
collinear classical ground states.
The other half of the spin-wave modes are the ordinary modes, which 
can be viewed as diamond lattice eigenmodes of Eq.~(\ref{eq:eig_tet}).
Of these bands, the \emph{optical ordinary modes} 
possess non-zero frequency throughout the Brillouin zone, 
and the \emph{acoustic ordinary modes}
have non-zero energy in most of the Brillouin
zone, but vanish along the major axes in reciprocal space
(see, for example, the solid lines in Fig.~\ref{fig:coplanar}).
We find that the number of acoustic zero modes does depend on the particular
classical ground state, and the zero modes in these bands are the divergent
modes, i.e. modes with divergent fluctuations 
$\langle \sigma^l_\qvec \sigma^m_{-\qvec} \rangle$.

Why are the divergent modes restricted to \emph{divergence lines}
in $\qvec$ space? 
Since the divergent modes are bounded, at most, in one direction,
the Fourier space basis is constructed by taking linear combinations of 
bounded planar modes, which are localized modes along the (major) axis
normal to the plane. 
We label the (100) modes in such a basis by $\{\wvec_m^x\}$, where
$x$ is a (real space) coordinate along the normal axis, and the index
$m$ reflects that there may be several types of plane modes in each direction
[for example, there could be four different types of square
divergent modes of the form shown in Fig~\ref{fig:div}b].
Fourier space divergent modes can thus be constructed by linear combinations
\begin{equation}
\wvec_m^{(q_x,0,0)} = \sum_x \wvec_m^{x} e^{i q_x x},,
\end{equation}
and similarly for $y$, $z$.
These linear combinations can be taken at any $\qvec$.
value along the normal axis, and the number of divergent modes along each
axis is equal to the number of values that the index $m$ can take.
The conclusion we can draw from this will be important later on
(in Sec.~\ref{sec:entropy}): the rank of the divergent mode space is of order
$N_s^{1/3}$.

If we look at the acoustic energy bands to which the divergent modes belong, 
moving away from the divergence lines, in $\qvec$ space, we find
(Fig.~\ref{fig:coplanar}) that the
energy increases linearly with $\qvec_\perp$, the component of $\qvec$ 
perpendicular to the divergence line.
The dispersion of the acoustic modes can be easily found analytically
by solving Eq.~(\ref{eq:eig_tet}) for small deviations
away from a planar divergent mode
(e.g. the one depicted in Fig.~\ref{fig:div}b),
with $q_\perp$ restricted to be \emph{within} the plane.

The singular spin fluctuations along lines in $\qvec$ space would produce sharp
features in the structure factor $S(\qvec)$, that could be measured in elastic
neutron diffraction experiments.
This is because the structure factor is proportional to the spin-spin
correlation
$\langle \mathbf{S}^\perp_\qvec \cdot \mathbf{S}^\perp_{-\qvec} \rangle$,
where $\mathbf{S}^\perp$ is the component of the spin transverse to
the scattering wave vector $\qvec$.
To lowest order in $S$, only $\sigma^x$ and $\sigma^y$ contribute to the
correlations, and one obtains
\begin{equation}
S(\qvec)\propto \sum_{\langle lm \rangle}
\langle \sigma_\qvec^l \sigma_{-\qvec}^m \rangle \,.
\end{equation}
Thus, the structure factor should have sharp features for any $\qvec$
along the major lattice axes.  

Furthermore, since the zero-point energy of the 
present harmonic theory will be shown to have degenerate
ground states (see Sec.~\ref{sec:gauge} and Ref.~\onlinecite{clh_harmonic}), 
anharmonic corrections to the harmonic energy determine the ground
state selection~\cite{uh_quartic}.
It turns out that the divergent modes become decisive in
calculating the anharmonic energy $\ham_\mathrm{quart}$.
The anharmonic spin-wave interaction would also serve to cut off the singularity
of the fluctuations.

In Sec.~\ref{sec:entropy}, we shall show that the divergent modes also provide
a useful basis for constructing and \emph{counting} gauge-like transformations
that relate the various degenerate ground states.

\subsection{Gauge-like symmetry}
\label{sec:gauge}

Upon examination of Eq.~(\ref{eq:eig_tet}),
it becomes apparent that the harmonic energy $E_{\harm}$,
of Eq.~(\ref{eq:zeropoint}),
is invariant under a $Z_2$ gauge-like transformation that changes the sign
of some tetrahedra spin deviations~\cite{clh_harmonic}
\begin{equation} \label{eq:gauge}
\eta_i \to \tau(\alpha) \tau(\beta) \eta_i\,,
\end{equation}
where 
$\tau(\alpha),\tau(\beta) \in \pm 1$, and $\alpha$, $\beta$ are the two
tetrahedra that share site $i$.
While this is an exact gauge symmetry of the projected (diamond lattice)
Hamiltonian (related to~(\ref{eq:eig_tet})), it is not a physical
gauge invariance, since the transformation must be
carried out in a way that conserves the tetrahedron rule,
i.e. does not take us out of the classical ground state manifold.
Furthermore, these transformations relate physically distinct states.
We will henceforth refer by \emph{transformation} to ``allowed''
gauge-like transformations that conserve the classical tetrahedron rule.

\begin{figure}[!h]
\resizebox{!}{5cm}{\includegraphics{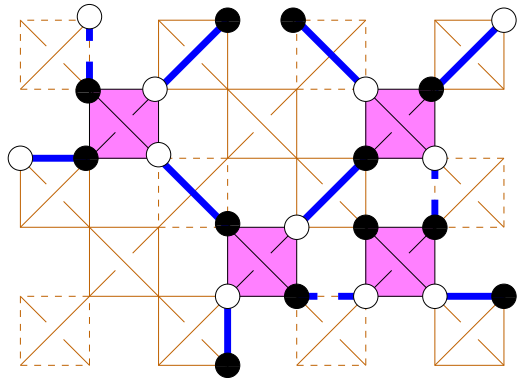}}
\caption{\footnotesize (Color online) (001) projection of a  portion of an 
even transformation.
The shaded tetrahedra, on the even sublattice of the diamond lattice,
are being flipped. Each flipped 
tetrahedron must be connected by $4$ satisfied (highlighted) bonds to
neighboring tetrahedra in the same sublattice.
Up (down) spins are signified by closed (open) circles.
Dashed lines connect one layer to another in the 
three-dimensional structure. \label{fig:gauge}}
\end{figure}

Fig.~\ref{fig:gauge} shows an example of a transformation that includes 
flipping some of the even tetrahedra only (\emph{even transformation}).
There are two transformations that can each be 
viewed either as an odd or as an even transformation:
the identity transformation ($\tau(\alpha)\!=\!1$ for any $\alpha$)
the global spin flip ($\tau(\alpha)\!=\!-1$ for any even $\alpha$
and $\tau(\alpha)\!=\!+1$ for any odd $\alpha$, or vice-versa).

For any even transformation $\tau_\even$, we can define its reverse (even)
transformation by 
\begin{equation}
\overline{\tau}_\even (\alpha) = -\tau_\even(\alpha) \,,
\forall \mbox{ even } \alpha \,,
\end{equation}
and similarly for odd transformations.
Examining Eq.~(\ref{eq:gauge}) we find that there are actually two ways of 
expressing any transformation as an product of an even transformation
and an odd transformation because
\begin{equation}
\overline{\tau}_\even \otimes \overline{\tau}_\odd \equiv
\tau_\even \otimes \tau_\odd \,,
\end{equation}
Even and odd transformations commute, in the sense that applying an even (odd)
transformation does not affect the set of allowed odd (even) transformations.
Thus we find that for any reference state, the number of possible gauge
transformations $\NN_{G}^{\rref}$ satisfies
\begin{equation} \label{eq:ngauge}
\NN_{G}^{\rref}= \half
\NN_{\even}^{\rref} \times
\NN_{\odd}^{\rref}  \,,
\end{equation}
where $\NN_{\even}^{\rref}$ and $\NN_{\odd}^{\rref}$ are the number
of even and odd transformations for the reference state, respectively,
including both the identity transformation and the overall spin flip.

If we take a particular classical ground state, and apply a gauge transformation
(as depicted in Fig.~\ref{fig:gauge}) to it, it is easy to see graphically what
would happen to the divergent modes (as shown in Fig.~\ref{fig:div}): 
If a tetrahedron marked by ``$+$'' in Fig~\ref{fig:div} overlaps the support of
the gauge transformation, it turns into a ``$-$'' and vise versa.
More formally, if we start from a given state, in which there is
a divergent mode $\uvec_d$,
a transformation $\boldsymbol{\tau}$
(such that $\tau_\alpha \!=\! \pm 1$) results in the new state with
a divergent mode
\begin{equation}
u(\alpha) \to \tau_\alpha u(\alpha) \,,
\end{equation}
for each $\alpha$.
Otherwise, the number and spatial support of the divergent modes
is gauge-invariant.
On the other hand, one finds that states that are not related by gauge
transformations, and therefore have different energies, generally have a
different number of divergent modes.

\section{Zero-point energy and effective Hamiltonian} \label{sec:zeropoint}

In this section, we study the zero-point energy of the harmonic
Hamiltonian~(\ref{eq:coll_harm_mat}).
First, in Sec.~\ref{sec:heff}, we write the energy as an expansion in real-space
paths, and use this expansion to produce an effective Hamiltonian in terms
of Ising like fluxes through the diamond lattice loops.
Next, in Sec.~\ref{sec:numerical}, we numerically diagonalize the Hamiltonian
to calculate the zero-point energy for various classical ground states.
We compare the results to the effective Hamiltonian and find that they agree
well, and predict the same ground state family. Finally, in 
Sec.~\ref{sec:entropy} we find an upper bound for the number of harmonic ground
states using a correspondence between gauge transformations and divergent modes.

\subsection{Effective Hamiltonian} \label{sec:heff}

As the zero-point energy is gauge invariant,
it can only depend on the gauge invariant 
combinations of $\{\eta_i\}$, which are products of the Ising variables
around loops.
These can be viewed as $Z_2$ flux through the plaquettes of the diamond
lattice~\cite{capped_kagome}.
In this section, we find an effective Hamiltonian in terms of these new degrees
of freedom, using a real space analytic loop expansion in the spirit 
of the loop expansions used in the large-$N$ calculations of
Refs.~\onlinecite{uh_LN,tchern_T_LN,tchern_LN}.

The harmonic energy of
Eqs.~(\ref{eq:zeropoint}),~(\ref{eq:coll_harm_mat}, and~(\ref{eq:heis_mat})
can be written as
\begin{equation} \label{eq:trace}
E_\harm = S \tr (\frac{1}{4} \muvec^2 )^{1/2}-SN_s \,,
\end{equation}
where $\muvec=\Wvec \etvec \Wvec^T$
is the $N_s/2 \times N_s/2$ matrix, whose indices are diamond lattice sites
\begin{equation} \label{eq:mu_el}
\mu_{\alpha \beta} = \sum_i W_{\alpha i}  W_{\beta i} \eta_i \,.
\end{equation}
For any collinear classical ground state,
the diagonal part of $\muvec$ vanishes, elements connecting 
diamond nearest neighbors are equal to $\pm 1$ all other elements are $0$. 
Therefore, the non-diagonal elements of
$\muvec^2$ connects between the same diamond sublattice, i.e. between
FCC nearest neighbors.
\begin{equation} \label{eq:mu2}
(\mu^2)_{\alpha \beta} = \left\{
\begin{array}{ll}
4 & \alpha=\beta\\
\eta_{\alpha \beta} & \alpha\,, \beta \mbox{ next nearest neighbors}\\
0 & \mbox{otherwise}
\end{array} \right. \,,
\end{equation}
where $\eta_{\alpha \beta}\!\equiv\! \eta_i \eta_j$ and $(ij)$ is the
(pyrochlore) bond connecting
$\alpha$ and $\beta$.
Thus, we could formally Taylor-expand the square root in
Eq.~(\ref{eq:trace}) about unity.
In order to assure convergence of the expansion, as will be discussed later,
we generalize this and expand about $A \openone$, where $A$ is an arbitrary
disposable parameter.
\begin{eqnarray} \label{eq:expand}
E_\harm&=&S \tr (A \openone + (\frac{\muvec^2}{4}-A \openone))^{1/2} -SN_s
 \\ &=&
S\sqrt{A} \sum_{n=0} C_n \tr(\frac{\muvec^2}{A} - 4 \openone)^n -SN_s 
\nonumber \\&=&
S\sqrt{A} \sum_{n=0} C_n \sum_{k=0}^n  \frac{(-4)^{n-k}}{A^k} \comb{n}{k}
\tr \muvec^{2k} -SN_s \,,\nonumber
\end{eqnarray}
with the coefficients
\begin{equation}
C_0 = 1 \,, \qquad C_n = (-1)^{n+1} \frac{(2n-3)!!}{8^n n!} \,,\mbox{for } 
n>0\,.
\end{equation}
$\tr{\muvec^{2k}}$ is a sum over all of the diagonal terms of
$\muvec^{2k}$,
i.e. a sum over products of $\muvec_{\alpha \beta}$ along all of the closed
paths of length $2k$ in the \emph{diamond} lattice.
Here a ``closed path'' is any walk on the diamond lattice that starts and ends at the same site.
This expansion involves constant terms that are independent of the sign of 
$\mu_{\alpha \beta}$ as well as terms that do depend on particular spin
configurations.
For example, any path of length $2k$, such that each step in one direction is
later retraced backwards (a \emph{self-retracing path}),
will contribute $1$ to $\tr \mu^{2k}$.
On the other hand, paths involving \emph{loops} on the lattice could contribute
either $+1$ or $-1$ depending on the spin directions.
Thus, we can re-sum Eq~(\ref{eq:expand}) to obtain an effective
Hamiltonian
\begin{equation} \label{eq:heff}
E_\harm^\eff = E_0 + K_{6} \Phi_{6} + K_{8} \Phi_{8}  +
\sum_{s(10)} K_{10,s} \Phi_{10,s}  
+\ldots \,,
\end{equation}
where $K_{2l}$, $K_{2l,s}$ are constants, which we calculate below, and
$\Phi_{2l}$ ($\Phi_{2l,s}$) are sums over all \emph{loops} of length $2l$ (and
type $s$).
The index $s$ is to differentiate between different types of loops of length
$2l$ that are not related to each other by lattice symmetries.
In our case, we do not need the index $s$ for loops of length $6$ or $8$, since
there is just one type of each.
On the other hand, there are three different type of loops of length $10$, 
and therefore there are three different $2l\!=\!10$ terms. 
Since we will explicitly deal with just the first three terms in 
Eq.~(\ref{eq:heff}), we shall omit the index $s$ from now on.
In fact, in the approximation that we present below, we will assume that 
all of the loops of \emph{any} length $2l$ have the same coefficient $K_{2l}$.

By our definition of loops, $\Phi_{2l}$
can be expressed either in terms of $2l$ diamond lattice sites 
along the loop $(\alpha_1,\alpha_2,\ldots,\alpha_{2k})$,
or in terms of $2l$ \emph{pyrochlore} lattice sites
$(i_1,i_2,\ldots,i_{2l})$
\begin{eqnarray} \label{eq:Phi}
\Phi_{2l} &\equiv&
\sum_{(\alpha_1,\alpha_2,...,\alpha_{2l})} \mu_{\alpha_1 \alpha_2}
\mu_{\alpha_2 \alpha_3} \cdots \mu_{\alpha_{2l} \alpha_1} \nonumber \\ &=&
\sum_{(i_1,i_2,...,i_{2l})}
\eta_{i_1} \eta_{i_2} \cdots \eta_{i_{2l}}
\,.
\end{eqnarray}
Note that in general, paths that we consider in the calculation 
(and that are not simple loops) should only be
viewed as paths in the diamond lattice.

\subsubsection{Bethe lattice harmonic energy} \label{bethe}

Before we evaluate the coefficients in Eq.~(\ref{eq:heff}) for the diamond
lattice, we shall, as an exercise, 
consider the simpler case of a coordination $z\!=\!4$ Bethe lattice.
In order for this problem to be analogous to ours, we assume that the number of
sites is $N_B\!=\!N_s/2$ and that each bond
$(\alpha \beta)$ in the lattice is assigned an Ising variable
$\mu_{\alpha \beta}\!=\!\pm 1$.
In this case, since each bond included in any closed path along the lattice is
revisited an even number of times,
and since $\mu_{\alpha \beta}^2\!=\!1$ for any
bond in the lattice, then each closed path of length $2k$ contributes 
$1$ to $\tr \muvec^{2k}$.
Thus all bond configurations in the Bethe lattice would have the same energy.

\begin{figure}
\resizebox{\columnwidth}{!}{\includegraphics{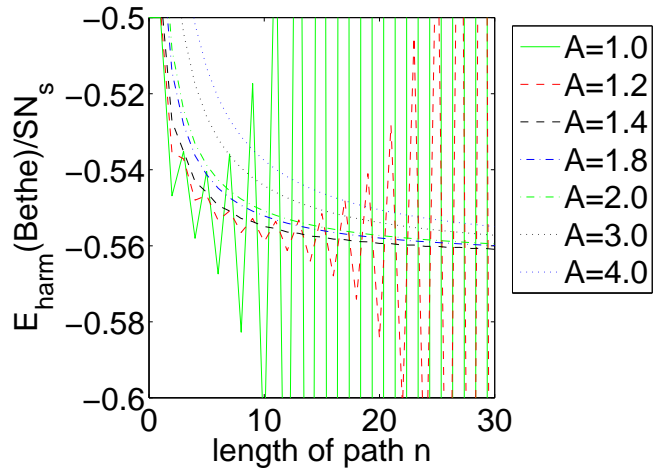}}
\caption{\label{fig:const} \footnotesize (Color online)
The analytically calculated constant term $E_0$, in the Bethe lattice 
approximation
as a function of the maximum path length considered, for various values of $A$.
We find that the sum converges for $A\!\agt\!1.4$.}
\end{figure}

\begin{figure}
\resizebox{\columnwidth}{!}{\includegraphics{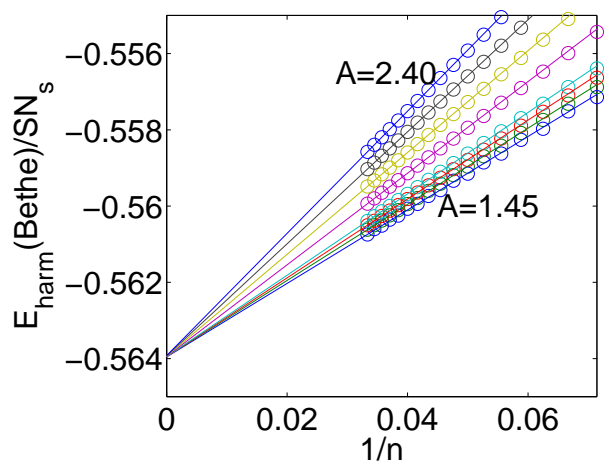}}
\caption{\label{fig:const_extra} \footnotesize (Color online)
Extrapolation of the energy
calculated for paths of length $n\!\le \! 30$ to $n\!\to\!\infty$,
using the linear dependence of $E_0$ on $1/n$. The calculated energy is shown
in open circles and the extrapolated inverse linear dependence in lines.
We see that the extrapolated results is nearly independent of our choice of 
$A\!>\!1.4$.  }
\end{figure}

Calculating the Bethe lattice energy for a given path length $2k$ turns out
to be a matter of \emph{enumerating} the closed paths on the Bethe lattice,
which can be done \emph{exactly} using simple combinatorics
(see App.~\ref{app:bethe}).
The sum~(\ref{eq:expand}) does not converge, as we consider longer and longer
paths, for the trivial choice of $A\!=\!1$, but converges well for $A \agt 1.4$
(see Fig.~\ref{fig:const}).
In the thermodynamic limit, we obtain (see Eq.~(\ref{eq:ebethe}))
\begin{equation} \label{eq:ebethe_num}
E_\harm(\mathrm{Bethe})=E_0=-0.5640 S N_s
\end{equation}
This value was obtained from Eq.~(\ref{eq:expand}), cutting it off at $n\!=\!30$
and extrapolating to $n\!\to\!\infty$ (see Fig.~\ref{fig:const_extra}).

\subsubsection{Bethe lattice approximation for the constant term}
\label{sec:const}

What can we learn from the Bethe lattice exercise about the diamond lattice
effective energy?
It turns out that actually the coordination-$4$ Bethe lattice calculation
provides a very good approximation for the constant term $E_0$.
There is a one-to-one correspondence between the Bethe lattice paths,
and self-retracing diamond lattice paths, which are the
paths that contribute to $E_0$.
Conversely, the product of $\mu_{\alpha \beta}$ along a
path that goes around a loop depends on the particular classical ground state,
and therefore contributes to the term in Eq.~(\ref{eq:heff}) corresponding to
that loop and not to $E_0$.

\begin{figure}
\resizebox{!}{3.5cm}{\includegraphics{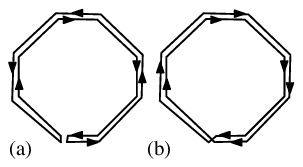}}
\caption{\label{fig:double_loop}\footnotesize 
Repeated loops that contribute constant terms: (a) Circled back and forth,
and accounted for by the Bethe lattice enumeration (self-intersecting paths),
and (b) Repeated in the same direction, not accounted for in our calculation.}
\end{figure}

There is only one type of path that contributes to the constant term, but was
omitted in the Bette lattice approximation:
If a loop is repeated twice (or any even number of times),
in the same direction, as in
Fig.~\ref{fig:double_loop}b, it \emph{does} contribute a 
constant term since each $\mu_{\alpha \beta}^2=1$ for each link along the loop.
A simple argument can be given to show that the constant terms 
involving repeated loops are negligible compared to the Bethe lattice terms.
First, we realize that the contribution of paths involving repeated loops
is exactly the same as the contribution of paths
that have already been counted by the Bethe lattice enumeration: self-retracing
paths that go around a loop,  and then return, in the opposite direction
(see Fig.~\ref{fig:double_loop}a). 
We call these \emph{self-intersecting} self-retracing paths since they 
include retraced diamond lattice loops. 
Since they still correspond to a Bethe lattice paths, 
they can be readily enumerated.

In order to get an idea for the number of such self-intersecting
paths, consider the smallest loop, a hexagon.
The number of paths of length $2k\!=\!12$ starting
from a particular point is (by App.~\ref{app:bethe}), $f_{6}\!=\!195352$, while
the number of hexagons touching that point is only $12$, accounting for $24$
intersecting paths of length $2k \le 12$!
If we now consider the paths of length $14$ starting from the same point, then
there are less than $100$ involving a repeated hexagon,
while there are nearly $2$ million in total.
Essentially, the number of self-intersecting paths,
involving loops of length $2l$ is smaller by a factor greater than $(z-1)^{l}$
than the total number of paths of the same length.

The argument is reinforced a posteriori, by enumeration of paths involving 
loops, that we do below.
The lowest order correction to the constant term $E_0$
(from Eq.~(\ref{eq:ebethe_num}) due to repeated loops is 
of the same order as the coefficient for a loop of particular loop of length 
$12$ (circling twice around a hexagon), which we find to be of order $10^{-4}$.

\subsubsection{One loop terms} \label{sec:loop}
We now move on to calculate the coefficients $K_{2l}$ of the
non-constant terms that involve simple loops.
The prediction of Ref.~\onlinecite{clh_harmonic}
is that these terms decay with increased length of loops 
and therefore an effective Hamiltonian of the form~(\ref{eq:heff})
can be derived.

Consider a particular loop of length $2l$. We try to enumerate all of the 
closed paths of length $2k$
that involve this loop and \emph{no other loops}, i.e. all 
of the terms proportional to $\prod_{i=1}^{2l} \mu_{\alpha_i \alpha_{i+1}}$,
where $\alpha_i$ are the sites along the loop ($\alpha_{2l+1}\!\equiv\!\alpha_1$).

Since we allow for no additional loops, we assume that the path is
a \emph{decorated loop}, i.e. a loop with 
with self-retracing (Bethe lattice-like) paths emanating from some or all of
the sites on it.
In order for this description to be unique, we allow the self-retracing path
emanating from site $\alpha_i$ 
along the loop to include site $\alpha_{i-1}$, but not site $\alpha_{i+1}$. 
Thus, the first appearance of the bond $(i,i+1)$ is attributed to the loop,
and any
subsequent appearance must occur after going back from a site $j\!>\!i$, and is 
attributed to the self-retracing path belonging to $j$. 

App.~\ref{app:loop} describes the practical aspects of this calculation.
See Fig.~\ref{fig:hex_diag} for a diagrammatic description of the paths 
we enumerate.
The approximation neglects, as before, the contribution of repeated loops,
which is negligible, by the argument of Sec.~\ref{sec:const}.
This means that, within out approximation, all loops of length $2l$ have the 
same coefficient in Eq.~(\ref{eq:heff}).
Calculating the sums in Eq.~(\ref{eq:expand}) for $n\!\le\!30$ and
extrapolating to $n\!\to\!\infty$, we obtain the values
$K_6\!=\!0.0136 S$, $K_8\!=\!-0.0033 S$.

Thus we have evaluated the first three coefficients in the effective Hamiltonian
~(\ref{eq:heff}), and in the next section, we shall compare the effective energy
to numerical diagonalization.

\subsection{Numerical diagonalization}
\label{sec:numerical}

In order to be able to test our predictions numerically, we first constructed
a large number of classical ground states, using  a path flipping
algorithm ~\cite{huse} on a cubic unit cell of $128$ sites, with
periodic boundary conditions.
In each step of this algorithm we randomly select a loop of alternating spins
to obtain a new classical ground state. 
Considering the large classical degeneracy, we can construct a very large 
number of collinear classical ground states in this manner. 
In order to explore diverse regions of the Hilbert space,
we started the algorithm with various different states that we constructed by
hand.

\begin{figure}[!h]
\resizebox{!}{7cm}{\includegraphics{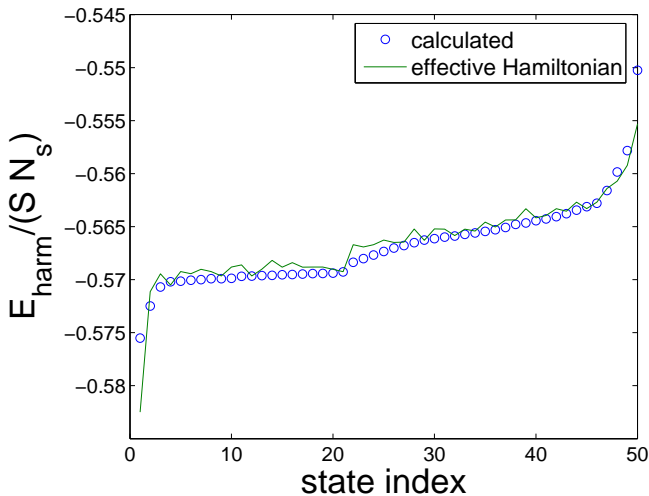}}
\caption{\footnotesize (Color online)
Calculated zero-point energy for 50 randomly generated classical ground 
states, 
compared to the Ising effective energy of Eq.~(\ref{eq:heff}), with $2l\le 8$.
State number $1$ in the plot is the $\pi$-flux state and state $50$ is the 
zero-flux state, in which all of the terms $\Phi_{2l}$ take the maximum
possible value. 
Notice that the effective Hamiltonian is not as good at calculating the energies
of these two extreme states, as
it does of calculating the energies of other states. This is because neglected
higher order terms in these states tend to add up in these states rather than
cancel out.
\label{fig:fit}}
\end{figure}

We have Fourier transformed the Hamiltonian~(\ref{eq:coll_harm_mat}), 
with a magnetic unit cell of $128$ sites, diagonalized for each $\qvec$ value,
and calculated the harmonic zero-point energy~(\ref{eq:zeropoint})
for fluctuation around a wide range of classical ground states.
We show the calculated $E_\harm$ for 50 sample states in Fig.~\ref{fig:fit}
\cite{footnote_2}
Our calculations verified that indeed gauge-equivalent states always have 
the same energy.

We also show in Fig,~\ref{fig:fit}
the effective energy $E_\harm^\eff$ for the same $50$ states, using
$K_6$ and $K_8$ calculated above.
The effective Hamiltonian, even with just 3 terms, proves
to do a remarkably good job of approximating the energy.
The root-mean-squared (RMS) error for the $50$ states shown in
Fig.~\ref{fig:fit}
is $1\!\times\!10^{-3}$, and it can be attributed to higher order terms
($2l\!>\!8$) in the expansion~(\ref{eq:heff}).
An independent numerical fit of the three constants $E_0$, $K_6$ and $K_8$
can lead to a slightly smaller RMS error, of the same order of magnitude,
but the coefficients are strongly dependent on the set of states used
in the fit.

\begin{figure}
(a) \resizebox{3.6cm}{!}{\includegraphics{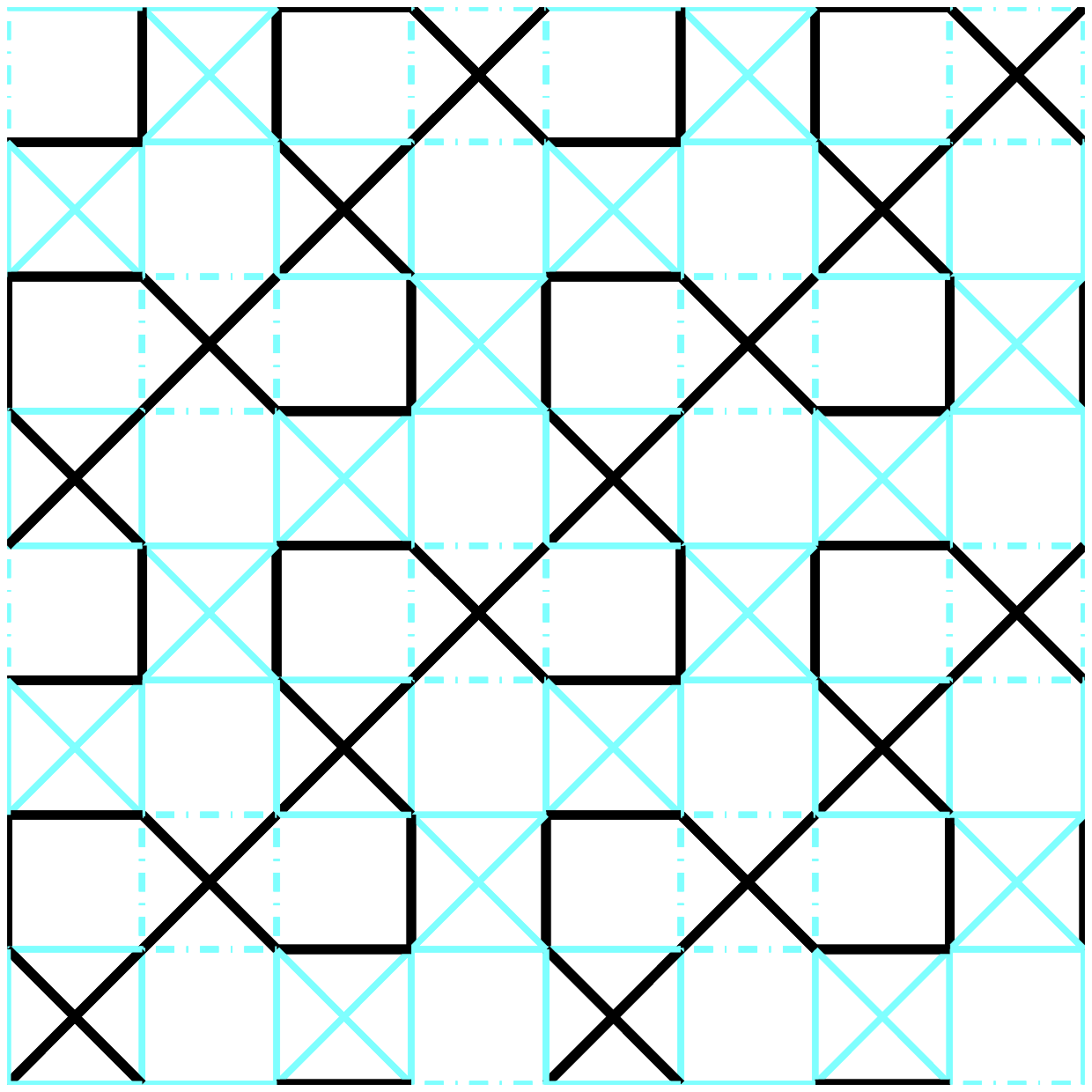}}
(b) \resizebox{3.6cm}{!}{\includegraphics{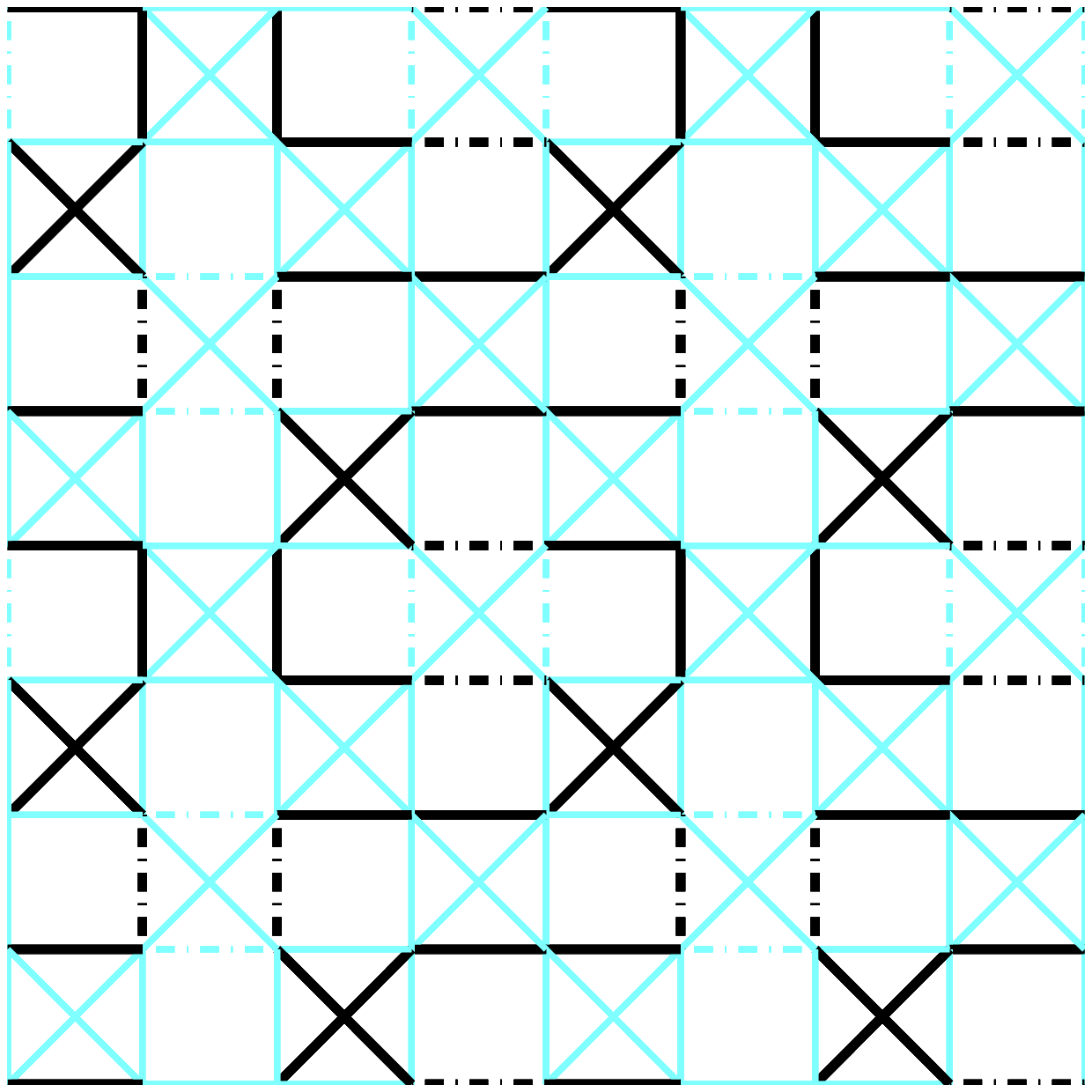}}
\newline
(c) \resizebox{3.6cm}{!}{\includegraphics{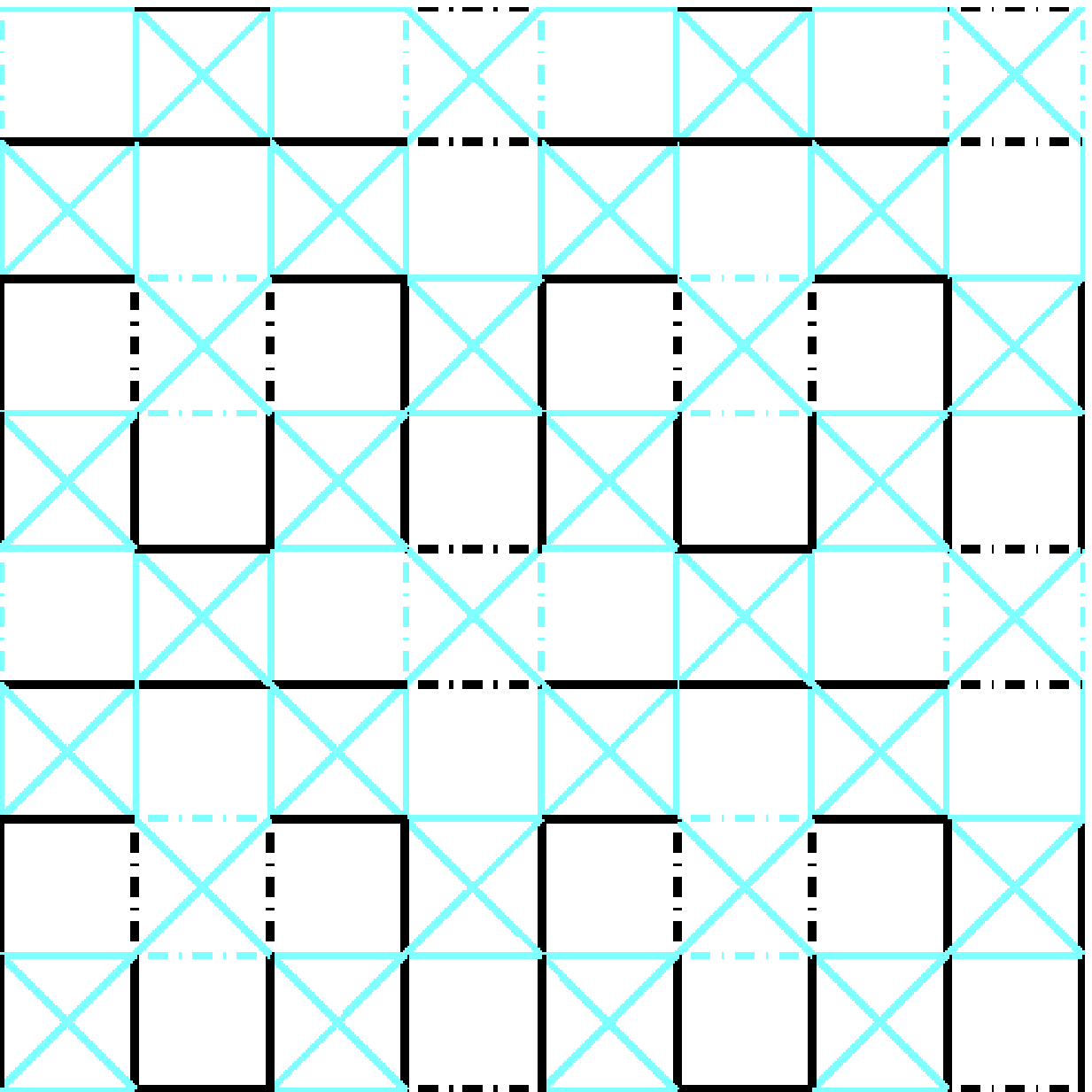}}
(d) \resizebox{3.6cm}{!}{\includegraphics{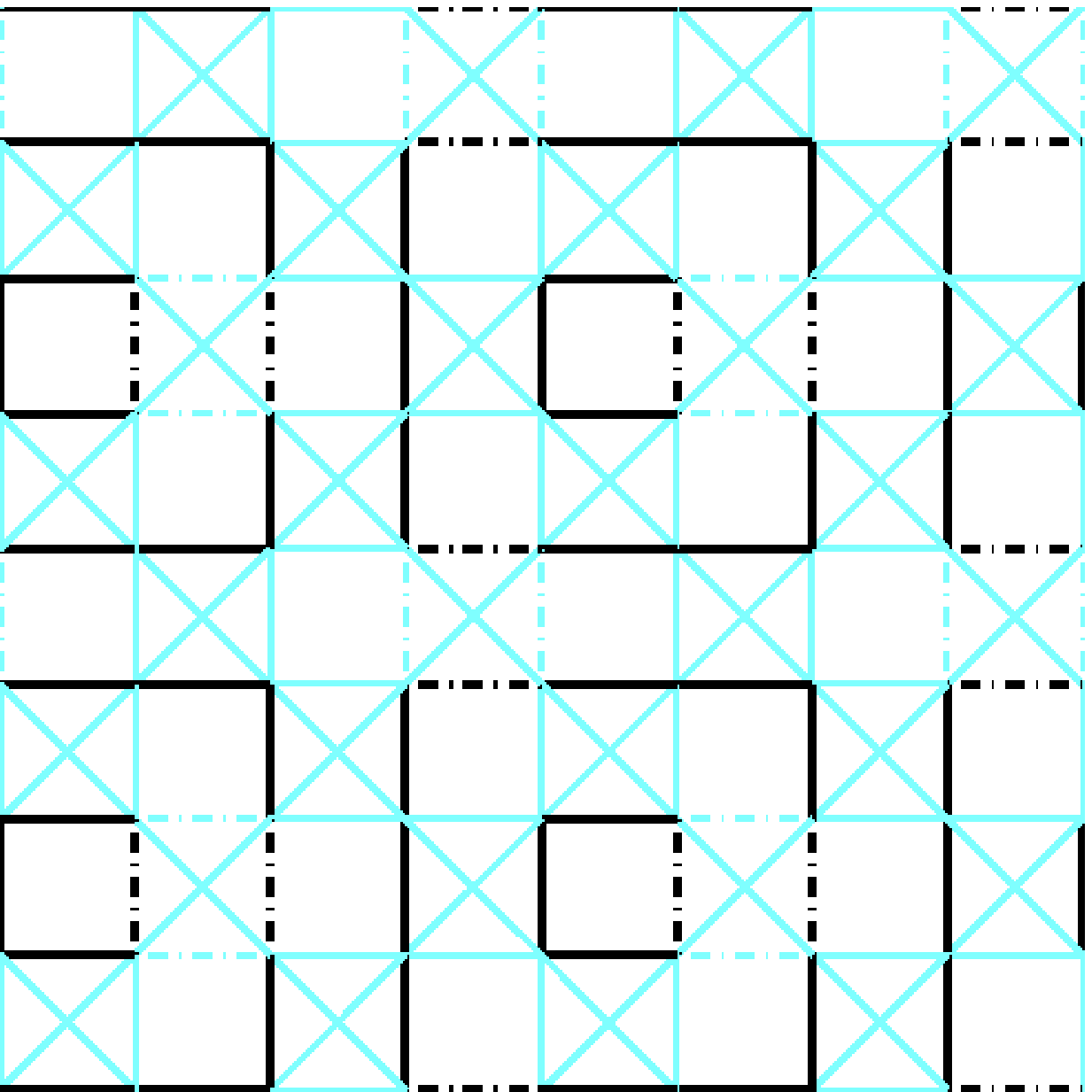}}
\caption{  \footnotesize (Color online)
(001) slices of some of the ground states with smallest magnetic unit cells.
For clarity, we only show the bond types,
which are (unlike the spin directions),
identical for all (001) slices, for these particular states,.
Ferromagnetic bonds are presented in dark and antiferromagnetic bonds are light.
Dashed lines connect this slice to adjacent slices.
\label{fig:states} }
\end{figure}

Based on the functional form of Eq.~(\ref{eq:heff}),
it has been speculated~\cite{clh_harmonic}, that the ground state 
manifold consists of all of the (gauge-equivalent) so-called
\emph{$\pi$-flux states}
(using the terminology of Ref.~\onlinecite{capped_kagome}), in which
\begin{equation} \label{eq:pi_flux}
\Pi_{\hexagon} \eta_i =-1 \,,
\end{equation}
for all hexagons.
We indeed find that there is a family of exactly degenerate ground states, 
satisfying Eq.~(\ref{eq:pi_flux}).
In Fig.~\ref{fig:states} we show some of the ground states. The smallest
magnetic unit cell that we obtain for a ground state has $16$ spins,
although if we consider bond variables rather than spins
(as in Fig.~\ref{fig:states}), the unit cell can be reduced to $8$ sites.
The highest energy, among collinear states, is obtained for the
\emph{$0$-flux states}, for which the spin directions have a positive product
around each hexagon.
Note that since the spin product around any loop can be written as a product of
hexagon fluxes, all states satisfying Eq.~(\ref{eq:pi_flux}) have the same
loop terms $\Phi_{2l}$, for any $l$, and are thus gauge-equivalent.
In general, if two states have the same hexagon product for every hexagon, they
are necessarily gauge-equivalent.

To search for ground states, we developed a computer
algorithm that randomly generates even or odd
gauge transformations starting from a particular state, and a unit cell
with periodic boundary conditions.
We start by flipping a random even tetrahedron.
In each subsequent step, we find an odd tetrahedron that violates the 
tetrahedron rule (i.e. has non-zero sum),
and randomly flip one of its (even) neighbors that can fix that violation.
This is repeated until there are no more violated tetrahedra.
A similar algorithm is employed to find odd transformations.

We performed an exhaustive search for ground states, satisfying
Eq.~(\ref{eq:pi_flux}), in the $128$ spin cubic magnetic unit cell
of linear dimension $L\!=\!2$.
We started with a particular $\pi$-flux ground state and 
randomly generated $10^7$ even and $10^7$ odd gauge
transformations and found $142$ unique transformations on each sublattice,
resulting, by Eq.~(\ref{eq:ngauge}),
in a total of $10082$ distinct states.
Only $24$ of these are unique with respect to lattice and spin-flipping 
symmetries.
Note that although the number of even gauge transformations and the
number of odd gauge transformations turn out to be the same for the harmonic
ground states, they need not be the same for other states. 
Fig.\ref{fig:states} shows the states with smallest unit cells.
By construction, all of these states are exactly degenerate to harmonic order
in spin-wave theory.
In future work, we shall explore the anharmonic selection among the harmonic
ground states~\cite{uh_quartic}.

\subsection{Ground state entropy} \label{sec:entropy}

In the preceding section we have found that there is a 
large family of exactly degenerate ground states.
Any two of these states are related by a gauge transformation of the type 
discussed in Sec.~\ref{sec:gauge}, and therefore,
in order to enumerate these states, we must find how many gauge transformations
one can perform on a given ground state, in an arbitrarily large system.
In Ref.~\onlinecite{clh_harmonic}, the number of ground states was
speculated to be of order $e^L$, where $L$ is the linear dimension of the
system size (as opposed to a classically extensive entropy).
However, that was only rigorously shown to be
a \emph{lower bound}, by explicitly constructing a set of layered ground
states in which each layer can independently be flipped.
Here, we aim to find an \emph{upper bound} for the number of ground states.

Comparing Figs.~\ref{fig:gauge} and~\ref{fig:div}, we see that there is a
close relation between even (odd) transformations and even (odd)
divergent modes. 
Any (diamond lattice)
divergent mode that is constructed with \emph{satisfied} antiferromagnetic
bonds only, describes a valid gauge transformation.
This is reminiscent of the relation that we saw in Sec.~\ref{sec:zero},
between generic zero modes 
of $\ham_\harm$ and transformations that keep $\ham_\cl$ invariant.
The relation 
between divergent modes and transformations, as illustrated above,
and our knowledge of the divergent modes allow us to
demonstrate that the entropy has an upper bound of order $L \ln L$.

Consider a particular reference harmonic ground state.
We assume that all of the ground states are related to each other by gauge
transformations.
Thus, any ground state is related to the reference state by a 
transformation, that can be almost uniquely expressed
as product of an even and an odd transformation.
The number of ground states $\NN_0$ is equal to the number of
transformations possible for the reference state $\NN_{G}^{\rref}$ 
(see Eq.~(\ref{eq:ngauge})).
\begin{equation}
\NN_0=
\NN_{G}^{\rref} \,.
\end{equation}
We have seen in Sec.~\ref{sec:band} that the number of independent divergent
modes is of order $L$.
In the $\pi$-flux harmonic ground states, we find that there are $24L$ 
independent divergent modes, where $L$ is measured in units of the 
underlying cubic lattice, half of which are even and half odd.
Consider an orthogonal basis of $12L$ real space planar even divergent modes,
i.e. $\{\uvec^m_\even\}$ are a set of unnormalized orthogonal vectors
living on the even diamond sublattice, satisfying Eqs.~(\ref{eq:u_orth})
and~(\ref{eq:div_ev}).
Any even transformation is associated with a divergent mode
$\uvec^G_\even$ also living on the even sublattice. 
We can write the transformation in terms of the basis
\begin{equation} \label{eq:gauge_sum}
\uvec^G_\even = \sum_{m=1}^{\NN_{\even}^{\rref}}
\frac{\uvec^G_\even \cdot \uvec^m_\even}{|\uvec^m_\even|^2} \uvec^m_\even\,.
\end{equation}
Each of the planar divergent modes has a support of $2L^2$ diamond lattice
sites, and since $\uvec^m_\even(\alpha)$ is $0$ or $\pm 1$  for any
$\alpha$ (Eq.~(\ref{eq:div_ev})), we find that $|\uvec^m_\even|^2= 2L^2$ 
Since $\uvec^G_\even(\alpha)$ is $0$ or $1$ for each $\alpha$,
the inner product $\uvec^G_\even \cdot \uvec^m_\even$ is also an
integer satisfying
\begin{equation}
|\uvec^G_\even \cdot \uvec^m_\even| \le |\uvec^m_\even|^2 \,.
\end{equation}
Therefore, each coefficient in Eq.~(\ref{eq:gauge_sum}) can take one of, at 
most, $4 L^2$ values, and since there are $12L$ basis
vectors $\uvec^m_\even$, the number of possible vectors $\uvec^G_\even$ is,
at most, $(2L)^{24L}$.
This is an upper bound on the number of even transformation, and similarly,
of odd transformations, as well.
By Eq.~(\ref{eq:ngauge}), we find
\begin{equation} \label{eq:degeneracy}
\NN_0\le \half  (2L)^{48L} \,.
\end{equation}
The entropy is defined as $\ln \NN_0$, and is, at most, of order $L \ln L$.
From Ref.~\onlinecite{clh_harmonic} we know that $\NN_0 \!>\! 2^4L$, and the
entropy is at least of order $L$.

This same bound on the order of the multiplicity applies to \emph{any} family of
gauge-equivalent Ising configurations, since we 
enumerated the possible gauge transformations on \emph{any}
given reference state (not necessarily a $\pi$-flux state),
which implies that the upper bound to the  number of states in \emph{any}
gauge-equivalent family is of the same order. 
However, while the order of magnitude
of the multiplicities of all energy levels
are the same, the coefficients in front of $L \ln L$ differ, because the
the number of independent divergent modes (which is always of order $L$) is
generally not the same for different gauge families.

\section{Non-collinear spins} \label{sec:noncoll}

We now move on to discuss the case of non-collinear classical ground states,
aiming to show that the energy of \emph{any} collinear ground state increases
upon rotating some of the spins out of collinearity.

\subsection{Collinear states are extrema of $E_{\harm}$ }

Consider first the case of a coplanar rotation, i.e.,
without loss of generality, some of the spins are
rotated such that the angle $\theta_i$ is neither $0$ nor $\pi$,
while $\phi_i$ remains $0$.
The elementary way of performing such a rotation is to rotate the spins in
an alternating (in $\eta_i$) loop, by $+\theta$ and $-\theta$ in an alternating 
fashion, where $\theta$ is a constant.
\begin{eqnarray}
\label{eq:el_rot}
\theta_i =0 &\to& \theta_i=\theta \,, \nonumber \\
\theta_i=\pi &\to& \theta_i=\pi-\theta \,.
\end{eqnarray}
%
Carrying through the derivation of Eq.~(\ref{eq:motion2}),
for the coplanar case, we find that the dynamical matrix elements change
\begin{equation} \label{eq:coplanar_dmatrix}
\eta_i \eta_k H_{ik} H_{kj} \to
\cos(\theta_i-\theta_k) H_{ik} H_{kj} \,.
\end{equation}
This means that in the expansion~(\ref{eq:expand}), $\muvec^2$ changes
from Eq.~(\ref{eq:mu2}) to 
\begin{equation}
(\mu^2)_{\alpha \beta} = \left\{
\begin{array}{ll}
4 & \alpha=\beta\\
\cos \theta_{\alpha \beta} & \alpha\,, \beta \mbox{ next nearest neighbors}\\
0 & \mbox{otherwise}
\end{array} \right. \,.
\end{equation}
Here $\theta_{\alpha \beta} \!\equiv\! \theta_i \!-\! \theta_j$, where $(ij)$ is
the (unique) pyrochlore bond a site $i\!\in\!\alpha$ and a site $j\!\in\!\beta$.

To see how the zero-point energy changes with coplanar deviations away from 
a collinear state, we take the derivative of the Hamiltonian~(\ref{eq:harm_mat})
with respect to $\theta_i$
\begin{equation}
\frac{\partial \ham_\harm}{\partial \theta_i}
= - \sum_j J_{ij} \sin(\theta_i-\theta_j) \sigma^\xx_i \sigma^\xx_j\,.
\end{equation}
This is clearly a zero operator for any collinear state.
Therefore
(i) all collinear states ($\theta_i\!-\!\theta_j\!\in\!\{0,\pi\}$)
are local extrema of $E_\harm(\{\theta_i\})$.
(ii) For small deviations of $\theta_i$ from collinearity,
the deviation of the energy from the value is quadratic:
\begin{equation}
E_\harm(\{\theta_i\})-E_\harm(0) = \OO(\{(\theta_i-\theta_j)^2\}) \,.
\end{equation}

\begin{figure}[!h] 
\resizebox{\columnwidth}{!}{\includegraphics{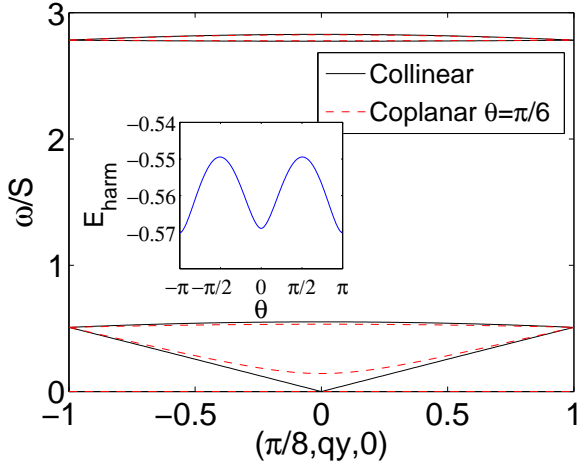}}
\caption{\label{fig:coplanar}
\footnotesize (Color online)
 Dispersion along a line in $\qvec$ space, for the collinear state shown in 
 Fig.~\ref{fig:states}a (solid lines), 
as well as the $\theta\!=\!\pi/6$ coplanar state (dashed lines),
showing that the divergent 
zero mode gains non-zero energy along the ($x$,$0$,$0$) line, when the state
is taken away from collinearity, while all other modes are virtually unchanged.
Insert: Zero-point energy of the same state with 
the spins along the ferromagnetic $x$-$y$ diagonal rotated by angle
$\pm \theta$.
The minimum energy is for the collinear states.
}
\end{figure}

We have shown that the collinear states are local extrema of $E_\harm$, 
in the space of coplanar classical ground states.
However, are they minima or maxima?
In order to find out, we take the second derivative of the Hamiltonian,
evaluated at a collinear state
\begin{eqnarray}
\left. \frac{\partial^2 \ham_\harm}{\partial \theta_i \partial \theta_j}
\right|_\mathrm{coll}
&=& J_{ij} \cos(\theta_i-\theta_j) (\sigma^\xx_i)^2
\nonumber \\ &&
 -\delta_{ij} \sum_k J_{ik} \cos(\theta_i-\theta_k) \sigma^\xx_i \sigma^\xx_k
\nonumber \\ &=& J_{ij} (\sigma^x_i)^2
 - \delta_{ij} \sum_k J_{ik} \sigma^x_i \sigma^x_k
 \,,
\end{eqnarray}
where we took $\cos(\theta_i-\theta_j)\!\to\!\eta_i \eta_j$ 
at the collinear state and used Eq.~(\ref{eq:sdev})
to transform to the $\sigma^x$ operators.

Now we can evaluate the elements of the Hessian matrix $\Mvec$
by taking the expectation value of this operator
\begin{eqnarray}
M_{ij}\equiv
\left. \frac{\partial^2 E_\harm}{\partial \theta_i \partial \theta_j}
\right|_\mathrm{coll}
= \left\{ \begin{array}{ll}
\langle (\sigma^x_i)^2 \rangle & i=j \\
\half J_{ij} \langle \sigma_i \sigma_j \rangle & \mbox{otherwise} 
\end{array}
\right. \,.
\end{eqnarray}
Here we used Eq.~(\ref{eq:fluct}) and the properties of the eigenmodes to 
simplify the expressions.
In order to show that the collinear states are local minima with respect to 
$\{\theta_i\}$, we would have to show that $\Mvec$ is positive definite.
Although we have not been able to shown this analytically,
we have found this to be true for a large number of periodic collinear states
that we have constructed (where we introduce a cutoff to the singularity 
of the fluctuations).

In order to examine the angle dependence in a general case, 
we note that the matrix elements $M_{ij}$ are dominated by the divergent
fluctuations of the divergent modes.
Therefore, the leading order change in $E_\harm$ is expected to come from the 
divergent modes.
We show below that the divergent modes' frequency is non-zero for coplanar
states, and therefore the zero point energy increases upon rotation from 
a collinear to a coplanar state.
\subsection{Spin-wave modes upon deviation from collinearity }
In order to better understand the origin of the quadratic (in $\theta$) 
energy change, we examine the eigenmodes of the new dynamical
matrix~(\ref{eq:coplanar_dmatrix}).
All of the $N_s/2$ generic zero modes of the collinear dynamical matrix are
zero modes of $\Rvec^\yy = \Hvec$ in the coplanar case
(using the local frame chosen in Sec.~\ref{sec:large_S}, see
Eq.~(\ref{eq:harm_mat}))
and therefore remain zero modes for \emph{any} coplanar spin arrangement.
One finds, however, that these generic zero modes acquire divergent
fluctuations in coplanar states,
because of the difference in stiffness between in-plane and out-of-plane
fluctuations.
This is similar to the case of the kagom\'e lattice, where all of the zero
modes of coplanar classical ground states have divergent fluctuation.

On the other hand, the divergent zero modes $\wvec_d$ of the collinear states,
become (non-divergent) nonzero modes when the spins are rotated out of
collinearity.
If a loop in a collinear state is rotated, as in~(\ref{eq:el_rot}),
by $\pm \theta$, 
we find that the divergent modes' frequency increases \emph{linearly} with
$\theta$. 
This rise in the divergent modes' zero-point energy is the reason that
each collinear classical ground state has \emph{lower} energy than nearby 
coplanar states.
However, after integration over the Brillouin zone, the rise in
total zero-point energy is quadratic in $\theta$.
The inset in Fig~\ref{fig:coplanar} shows a numerical example of this for the
state in Fig.~\ref{fig:states}a, where all of the spins along the
ferromagnetic line, in the (110) direction, are rotated by a small angle away 
from the axis of the other spins. Clearly, the lowest energy configurations
are collinear. In Fig.~\ref{fig:coplanar}, we see that the most significant 
difference in the dispersion between collinear and non-collinear states, is the
gap formed in some of the zero modes along the divergence lines.
The change in energy of the optical modes is negligible compare to this gap,
for all cases that we looked at.

Looking at further rotation of spins out of a \emph{coplanar} arrangement,
i.e. rotation of a loop or line by angles $\pm \phi$,
one finds that some of the zero modes gain non-zero frequency, proportional 
to $|\phi|$, as for the kagom\'e lattice~\cite{ritchey,vdelft},
and the energy increase is $\OO(\{\phi_i\})$.

While we have shown that collinear states are local minima of the energy
landscape, we have not ruled out the possibility that a non-collinear state 
would be a local (or even global) minimum.
Although we do not believe this to be the case, further work would be required
to prove so.

\section{Effective Hamiltonian for related models} \label{sec:other_models}

The loop expansion of Sec.~\ref{sec:heff}
can be easily be adapted to study similar models that support collinear 
ground state including the pyrochlore Heisenberg model with a large
applied magnetic field
and the Heisenberg model on closely related lattices, as we shall show in 
Secs.~\ref{sec:field} and~\ref{sec:other_lattices}, respectively.

For the purpose of determining the ground state manifold, we find it convenient
to recast the effective Hamiltonian of Eq.~(\ref{eq:heff}) as an Ising model
in the \emph{complementary lattice}.
This is a lattice composed of the centers of the shortest loops,
or \emph{plaquettes}, in the original lattice.
In the pyrochlore lattice, the complementary lattice sites are the centers
of the hexagons, and form another pyrochlore lattice. 
To each complementary lattice site $a$, we assign an Ising spin
$\tdeta_a$, equal to the product of the direct lattice sites around the
corresponding plaquette.
\begin{equation}
\tdeta_a \equiv \prod_{i \in a} \eta_i \,,
\end{equation}
where the product is on all sites in the (direct lattice) plaquette $a$.
Since any loop product can be written as a product of spin products around
plaquettes, the terms $\Phi_{2l}$ is Eq.~(\ref{eq:heff}) are now represented by 
simpler (at least for small $l$) expressions, in terms of the complementary 
lattice spins
\begin{eqnarray} \label{eq:comp_heff}
E_\harm^\eff &=& \EE_0 + S \BB \sum_{a} \tdeta_a +
S \JJ \sum_{\langle a b \rangle} \tdeta_a \tdeta_b \nonumber \\
&&+ S \JJ' \sum_{\llangle a b \rrangle} \tdeta_a \tdeta_b + 
S \JJ_3 \sum_{\triangle  a b c } \tdeta_a \tdeta_b \tdeta_c +\ldots \,,
\end{eqnarray}
where $\langle \cdots \rangle$ and $\llangle \cdots \rrangle$ represent nearest
neighbors and next-nearest neighbors on the complementary lattice, respectively,
and the $\triangle$ sum is over $3$-spin plaquettes.
Comparing to Eq.~(\ref{eq:heff}), we can identify, for the pyrochlore, 
$\EE_0\!\equiv\! E_0$, $\BB \!\equiv\! K_6/S$, $\JJ \!\equiv\! K_8/2S$,
$\JJ'\!\equiv\! K_{10}/S$, $\JJ_3\!=\!0$.
The $1/2$ factor in $\JJ$ stems from the fact that each $8$-loop in the
pyrochlore has two different representations as a product of two hexagons,
and $\JJ_3$ vanishes because in pyrochlore complementary lattice the product
of three spins of a tetrahedron is equal to the fourth spin in the same
tetrahedron. This is due to the dependence, in the direct lattice,
between the four hexagons in one super-tetrahedron
(see Fig.~\ref{fig:zero_modes}b).
Thus the $\JJ_3$ term is already accounted for in the ``field'' $\BB$ term.
Note that, in the pyrochlore, there are three different types of loops of length
$10$, and thus there will be $3$ different terms in Eq.~(\ref{eq:comp_heff})
that have a coefficient equal to $\JJ'$, within the decorated loop
approximation.

Writing the effective Hamiltonian in terms of the complementary lattice spins
is manifestly gauge invariant, since the complementary spins $\tdeta_a$ are not
modified by gauge transformations.
In the pyrochlore Heisenberg model, we found in Sec.~\ref{sec:zeropoint}
a ferromagnetic nearest neighbor interaction $\JJ$, with a positive field $\BB$,
resulting in a uniform complementary lattice ground state in which
$\tdeta_a\!=\!-1$ for all $a$ (the $\pi$-flux state).
This unique complementary lattice state corresponds to the family of
direct lattice ground states satisfying Eq.~(\ref{eq:pi_flux}).

\subsection{Non-zero magnetic field} \label{sec:field}

We now consider what happens to the loops expansion
when a magnetic field is applied to the system.
Since quantum fluctuations favor collinear spins, one generically expects
the magnetization to field curve to include plateaus at certain rational
values of the magnetization, corresponding to collinear spin
arrangements~\cite{ueda,penc+bergman,zhp_field}.
Thus, at certain values, the field will choose collinear states satisfying
\begin{equation} \label{eq:m_constr}
\sum_{i \in \alpha} \eta_i = M \,, \qquad \forall \mbox{ simplexes } \alpha\,,
\end{equation}
for some non-zero $M$.
These are the new classical ground states in this theory, and now the question
we ask is which of these are selected by harmonic quantum fluctuations.
Repeating the derivation of the loop expansion~(\ref{eq:expand}), we now find
that the diagonal elements of $\Hvec$ change, as now Eq.~(\ref{eq:heis_mat}) is 
modified to $\Hvec=\frac{1}{2} \Wvec^T \Wvec - \etvec M$.
Writing down the equations of motion in terms of the diamond lattice, as before,
we find that Eq.~(\ref{eq:trace}) is still valid, but that the
diagonal elements of $\muvec$ are all non-zero and equal to $-M$.
In order to remove these diagonal elements, we can define
\begin{equation}
\muvec_0 \equiv \muvec + M \openone \,,
\end{equation}
where $\muvec_0$ is equal to the $M\!=\!0$ value of $\muvec$ as defined in
Eq.~(\ref{eq:mu_el}), and, as before, it
only connects nearest neighbor tetrahedra. The loop expansion is now
\begin{eqnarray} \label{eq:h_expand}
E_\harm&=&S \tr \left(A \openone + \left(\frac{(\muvec_0-M \openone)^2}{4}-
A \openone\right)\right)^{1/2} \nonumber \\ &&- SN_s + |M| N_z S
\nonumber \\ &=&
S\sqrt{A} \sum_{n=0} C_n \tr\left(\frac{(\muvec_0^2-2M \muvec_0) +
(M^2-4A) \openone}{A}\right)^n \nonumber \\ &&-SN_s  + |M| N_z S
\nonumber \\&=&
S\sqrt{A} \sum_{n=0}^\infty
\frac{C_n}{A^n}\sum_{k=0}^n \sum_{j=0}^{n-k}  \tcomb{n}{k}{j}
\nonumber \\ &&\times 
 (M^2-4A)^{n-k-j} 
(-2M)^j \tr \muvec_0^{2k+j} \nonumber \\ &&-SN_s + |M| N_z S \,.
\end{eqnarray} 
Here $N_z$ is the number of zero modes of $\Wvec^T \Wvec$, i.e. $N_s/2$ for the
pyrochlore lattice.
Now the calculation goes as in Sec.~\ref{sec:heff},
noting that the trace is non-zero only for even $j$.
When one re-sums the terms in Eq.~(\ref{eq:h_expand}) to construct an effective 
Hamiltonian of the form~(\ref{eq:heff}), 
one finds that unlike the $M\!=\!0$ case,
where always $\mathrm{sign}(K_{2l}) = (-1)^{l+1}$, the signs of the 
expansion terms are no longer easy to predict.

Applying this calculation to the only non-trivial collinear case on the
pyrochlore lattice, i.e., $M\!=\!2$, and rearranging the terms in the
form~(\ref{eq:comp_heff}), we find (see Tab.~\ref{tab:summary}) that
$\BB\!<\!0$, and that the coefficient of the interaction terms, $\JJ$,
is two orders of magnitude smaller than the in effective field $\BB$.
Therefore the complementary lattice ground state is a uniform $\tdeta_a\!=\!1$ 
state, corresponding to the family of states with positive hexagon
products (the so-called \emph{zero-flux states}).

As in the zero magnetization case, studied in
Secs.~\ref{sec:model}--\ref{sec:zeropoint},
the harmonic energy is invariant under
any gauge transformation that flips the spins in a set of tetrahedra,
while conserving the constraint~(\ref{eq:m_constr}).
However, it is clearly much harder, in this case, to construct a
transformation in this way,
because a single tetrahedron flip violates the constraint not
just on the neighboring tetrahedra, but on the flipped tetrahedron itself 
as well.
Therefore one would expect the ground state degeneracy for this model to be 
smaller than for the $M\!=\!0$ case.

\begin{figure}
\resizebox{8cm}{!}{\includegraphics{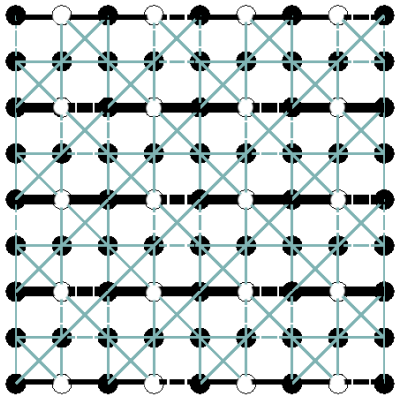}}
\caption{  \footnotesize (Color online)
(001) projection of a pyrochlore lattice $M\!=\!2$ $0$-flux ground state.
Each tetrahedron has $3$ up spins (filled circles) and one down spin (empty 
circle).
This is the ground state with the smallest possible ($4$ site) magnetic unit
cell.
The dark lines show $x$-$z$ planes (coming out of the page), that can each 
by flipped independently to obtain a valid $0$-flux $M\!=\!2$ state. 
Since there are $\OO(L)$ such planes, the ground state entropy is 
at least of order $L$.
\label{fig:pyro_m2_gs} }
\end{figure}

It is easy to show that still,
the ground state entropy is \emph{at least} of order
$L$, by observing that in the simplest ground state,
with a $4$ site unit cell, on can construct gauge
transformations by independently flipping some of $\OO(L)$ 
parallel planes, each composed e.g. of parallel antiferromagnetic lines in 
the $x$-$z$ plane (see Fig.~\ref{fig:pyro_m2_gs}).
The keys to these transformations being valid are:
(i) The spin-product around any hexagon (and therefore any loop) in the lattice
is not affected by it.
(ii) Since antiferromagnetic lines are flipped, the spin sum on each tetrahedron
remains the same.

\subsection{Other lattices} \label{sec:other_lattices}

The calculation of the loop expansion is actually quite general and can be
applied on any lattice composed of corner-sharing simplexes,
as long as the classical ground state is collinear,
and the (zero field) Hamiltonian has the form~(\ref{eq:heis_mat}).
We have implicitly assumed that the \emph{simplex lattice},
formed by centers of simplexes, is 
bipartite, although we could easily modify the calculation to take care of a
more general case.
Given these assumptions, the only lattice information relevant to our 
calculations is the coordination $z$ of the simplex lattice and the
lengths of the various loops in the same lattice.
In most cases, the coordination of the simplex lattice is equal to the number of
sites in a single simplex, e.g. $z=4$ for the pyrochlore,
or $3$ for the kagom\'e.
However the expansion works equally well for cases where not every lattice
site is shared by two simplexes, as in the \emph{capped kagom\'e} model below,
in which case $z$ is smaller than the number of sites in a simplex.

In general, the ratio $|K_{2l+2}/K_{2l}|$ is expected to increase
and becomes closer to $1$ for smaller values of $z$,
meaning that we must take longer and longer loops into consideration, to 
determine the ground state.
We find numerically that the convergence rate of the expansion~(\ref{eq:expand})
decreases as $z$ decreases, as does the value of $A$ at which convergence is
obtained.
The accuracy of our Bethe lattice approximation
is expected to be better for larger values of
$z$, since there are relatively fewer uncounted paths. 
%

\subsubsection{Checkerboard lattice}
The checkerboard lattice is also often called the planar pyrochlore, and is
a two-dimensional projection of the pyrochlore
lattice~\cite{lieb_check,canals_check}.
It is composed of ``tetrahedra'' (crossed squares)
whose centers form a square lattice (see, e.g. the (001) projections of
the pyrochlore in
Figs.~\ref{fig:zero_modes}a,\ref{fig:div},\ref{fig:gauge},\ref{fig:states}). 
Aside from the dimensionality, a major difference between the checkerboard
and the pyrochlore lattice is that in the checkerboard, not all of the
tetrahedron bonds are of the same length, and thus one would generally expect
coupling $J'\ne J$ along the square diagonals.
Nevertheless, to pursue the frustrated analog of the pyrochlore, we shall
consider here the case $J'\!=\!J$.

As in the diamond lattice, the coordination of the (square) simplex
lattice is $z\!=\!4$, 
and therefore the calculation of approximate loop expansion coefficients,
is identical to the pyrochlore calculation of Sec.~\ref{sec:heff}.
However, we should note that since the shortest loops are now of length $4$, 
the error in our Bethe lattice approximation is greater than in the pyrochlore
case, since the ``repeated loops'' that we ignore carry more significant weight.
Nevertheless, the ignored terms are still expected to be two orders of magnitude
smaller than the Bethe lattice terms.

In the checkerboard case, the complementary lattice is a square lattice composed
of the centers of the empty square plaquettes.
Now the effective field in Eq.~(\ref{eq:comp_heff}) is the 4-loop coefficient
$\BB\!=\!K_4/S\!<\!0$, which prefers $\tdeta_a\!>\!0$, or in other words,
$0$-flux order.
While the nearest neighbor complementary lattice coupling is 
antiferromagnetic $\JJ\!=\!K_6/S\!>\!0$ and competes with $\BB$,
it is not strong enough (see
Tab.~\ref{tab:summary}) to frustrate the uniform $0$-flux order.

The ground state entropy of the checkerboard lattice, has been shown in
Ref.~\onlinecite{tcher_check} to be of order $L$, by construction
of all of the possible even and odd gauge transformations on a particular 
reference $0$-flux state. 
A simple explanation for the degeneracy is that, given 
a line of spins, say in the $x$ direction, there are, at most, 
$2$ choices in the construction of an adjacent parallel line.
This is because there is one constraint on each tetrahedron ($2$ down-spins)
and one on each plaquette (an even number of down-spins).
Thus, starting from an arbitrary choice of one horizontal line ($2^L$ choices), 
there are $\le\!2^L$ ways of constructing the rest of the lattice.
The resultant entropy is $\OO(L)$.

Applying a magnetic field that induces the $M\!=\!2$ plateau, we find
(Tab.~\ref{tab:summary}), that the complementary lattice effective
Hamiltonian has $\BB\!>\!0$, $\JJ\!<\!0$, favoring the $\pi$-flux uniform state.
The ground state entropy in this case is also of order $L$.
To show this, we note that there is only one down-spin in each tetrahedron when
$M\!=\!2$, and that the spin product around each plaquette is $-1$.
This means that there must be exactly one down-spin around each plaquette.

We could use an argument similar to the one used in the  $M\!=\!0$ model, 
to find the entropy in this model.
A more elegant argument uses a $1$-to-$1$ correspondence between the
ground states
and \emph{complete} tiling the checkerboard lattice with squares of size
$2a\!\times\!2a$, where $a$ is the size of each plaquette (and ``tetrahedron'').
Here each square is centered on a down-spin and covers the two plaquettes and
two tetrahedra to which it belongs.
The entropy of such tilings is clearly of order $L$.


\subsubsection{``Capped kagom\'e''}
The kagom\'e lattice Heisenberg model has been one of the most
studied highly frustrated models.
The lattice is two dimensional, and is composed of corner sharing triangles, 
such that the centers of the triangles form a honeycomb simplex lattice. 
This model too is closely related to the pyrochlore, as a (111) projection of
the pyrochlore lattice contains kagom\'e planes sandwiched between triangular
planes, such that the triangular lattice sites ``cap'' the kagom\'e triangles
to form tetrahedra.

\begin{figure}
\resizebox{\columnwidth}{!}{\includegraphics{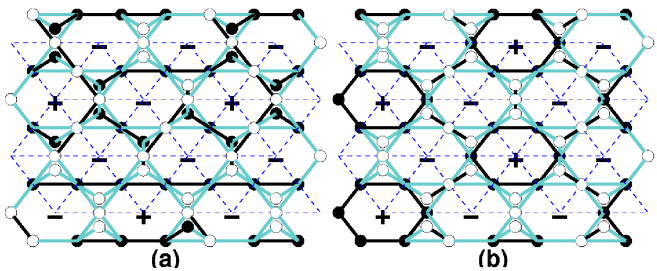}}
\caption{ \label{fig:capped_kag_gs}
\footnotesize (Color online)
(a) The $2\!\times\!2$ ground state of the ``capped kagom\'e'' lattice.
(b) The $\sqrt{3} \!\times\! \sqrt{3}$ state in the same lattice.
Up (down) spins are marked by filled (empty) circles, and
ferromagnetic (antiferromagnetic) lattice bonds are shown in dark (light).
The dashed lines represent the complementary lattice bonds, whose sites are 
marked $+$ and $-$.}
\end{figure}

We cannot apply our collinear loop expansion to 
the kagom\'e Heisenberg model,
with no applied field, because there are no collinear states that can satisfy
the zero triangle-sum rule. 
One way to consider a ``collinear kagom\'e'' is to look at a
\emph{capped kagom\'e} lattice
which consists of a kagom\'e, flanked by two triangular lattices, so that
each triangle turns into a tetrahedron. 
This model was studied by Tchernyshyov et. al.~\cite{capped_kagome}, who 
referred to it as a ``$[111]$ slice of pyrochlore''.
Those authors found that the ground state is one in which one out of every
$4$ hexagons has a positive spin product (the \emph{$2 \!\times\! 2$ state} 
shown in Fig.~\ref{fig:capped_kag_gs}a).
Surprisingly, the Hamiltonian for this model can 
be written in the form~(\ref{eq:heis_mat}),
and therefore we expect our loop expansion to work.
Furthermore, as long as there is no applied field, $\BB\!=\!K_6/S$ is positive, so applying our intuition based on the previously discussed models, we would
naively think that the ground state should be a $\pi$-flux state.

However, this is the one model that we have studied,
where the complementary lattice effective 
Hamiltonian has interaction terms strong enough to resist the $\BB$ field term.
Here, the complementary lattice nearest neighbor term corresponds to
loops of length $10$ and therefore $\JJ\!>\!0$ (there are no loops of length $8$
in the kagom\'e), and $\JJ'\!=\!K_{12}/S\!<\!0$.

In the $\pi$-flux state, all complementary lattice spins are $-1$, and thus
while the $\BB$ term in Eq.~(\ref{eq:comp_heff}) is optimized, all
of the nearest neighbor bonds ($3N_s^c$, which $N_s^c$ is the number of
complementary lattice sites) are violated, as well as all of the
$3$-site terms ($2N_s^c$).
On the other hand, the $2\!\times\!2$ state has only $N_s^c/4$ negative
spins, but half of the complementary lattice bonds and $3/4$ of the $3$-spin
terms are satisfied.
Applying the coefficients we obtained (from Tab.~\ref{tab:summary},
$\JJ/\BB \approx 0.14$), and including also the next order ($K_{12}$) term
$\JJ_3/\BB\approx -0.05$) to the expansion,
we find the energy per complementary-lattice site
\begin{equation} \label{eq:capped_kag_de}
\frac{E_\harm^{\pi\mathrm{-flux}}-E_\harm^{2\times 2}}
{S N^c_s} \approx -0.5 \BB+ 3 \JJ - 3 \JJ_3
\approx 0.0026>0 \,,
\end{equation}
and we find that the $2\!\times\! 2$ state is favored over the $\pi$-flux
state~\cite{capped_kagome}.
We provide this calculation as an illustration for the difficulty in 
using of the effective Hamiltonian in a model with an unusually
frustrated complementary lattice. 
In order to actually determine the ground state in this case, one must include
further terms in the effective Hamiltonian.
In fact, based only on the terms included in Eq.~(\ref{eq:capped_kag_de}), 
we would conclude, mistakenly, that the complementary lattice ground state
is the \emph{$\sqrt{3} \!\times\! \sqrt{3}$ state} (not to be confused with
the well-known coplanar kagom\'e ground state with the same ordering vector),
where $1/3$ of the spins are up, so that each triangular plaquette 
has 2 down spins and one up spin (see Fig.~\ref{fig:capped_kag_gs}b).
However, based on numerical diagonalization, we find that both the
$2\!\times\!2$ state and the $\pi$-flux state have, in fact, lower energy than
the $\sqrt{3}\!\times\!\sqrt{3}$ state, in agreement with
Ref.~\onlinecite{capped_kagome}.

Due to the large number of degrees of freedom in this model, arising from the
``free'' spins capping each triangle, there is an extensive number of
$2\!\times\!2$ ground states, as has been calculated in
Ref.~\onlinecite{capped_kagome}.

\subsubsection{Kagom\'e with applied field}
Another way of obtaining a ``collinear kagom\'e'' model,
is to apply a field strong enough to induce collinear states with $M\!=\!1$
on the (standard) kagom\'e lattice~\cite{zhp_field,zhit_field}.
Applying our expansion, we find $\BB\!<\!0$ and  $\JJ\!<\!0$
(see Tab.~\ref{tab:summary}), consistent with a uniform zero-flux ground state,
as we have indeed confirmed by numerically calculating the zero-point energy.

\begin{figure}
\resizebox{!}{4.0cm}{\includegraphics{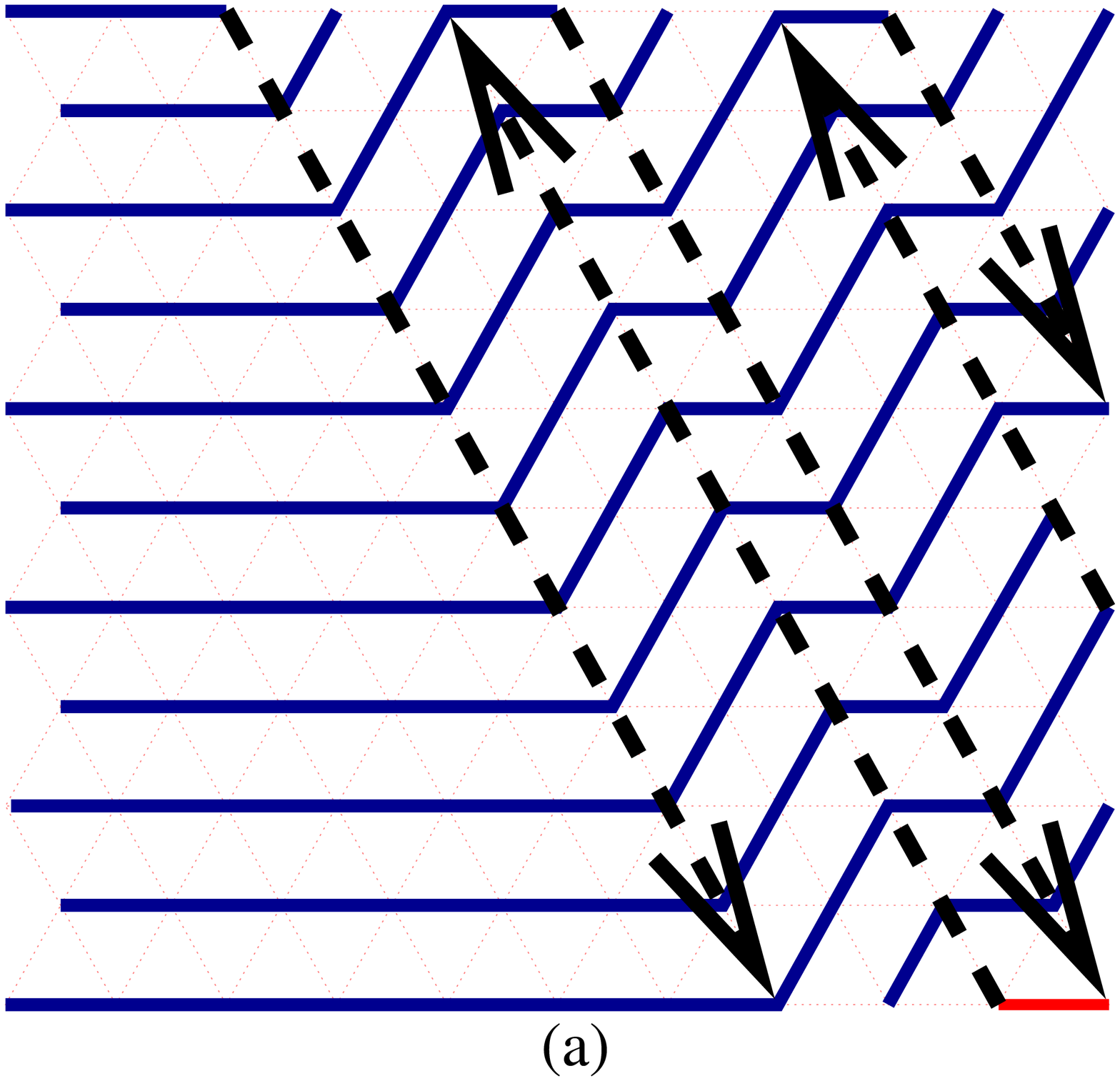}}
\resizebox{!}{4.0cm}{\includegraphics{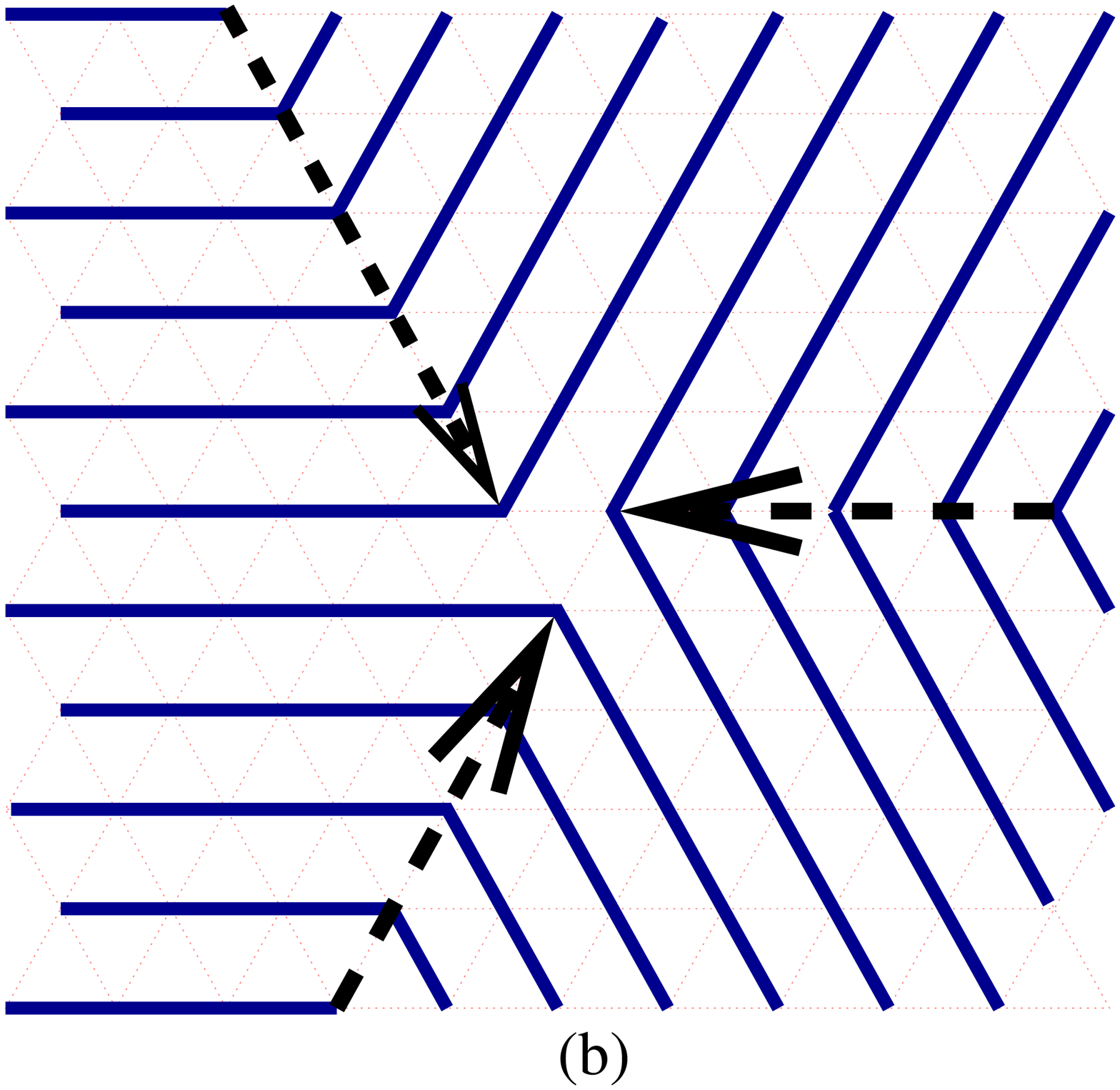}}
\caption{\label{fig:kag_m1_lines} (Color online)
The $M\!=\!1$ kagom\'e model ground states can be mapped $1$-to-$1$ to 
continuous lines on triangular lattice, such that a line can turn, at each 
site by $\pm 60^\circ$ or $0^\circ$. The ``kinks'' in each lines must lie
along directed defect lines (dashed lines).
There are two possible cases: (a) parallel defects alternating in directions.
(b) three lines meeting at a point defect (and no other kinks).}
\end{figure}

In the $M\!=\!1$ kagom\'e model, $0$-flux states are obtained by spin
arrangements in which there are exactly $2$ down-spins around each hexagon.
To find the ground state entropy, we map these to 
a dimer covering of the (triangular) complementary lattice, in which 
there are exactly $2$ dimers touching each complementary lattice site, 
and there is exactly $1$ dimer in each plaquette. 
These can be viewed as continuous lines on the triangular lattice, that 
can turn at each node by $\pm 60^\circ$ or $0^\circ$. 
Since there is no way of closing such lines into loops, one finds that they 
generally run in parallel, possibly turning by $60^\circ$ to form a 
``directed line defect''.
Three such line defects can meet and terminate at a point, as long as 
they form $120^\circ$ angles and are all directed into this ``point defect''.
Thus, each ground state either has exactly one point defect and three line 
defects coming into it (and no others), as in Fig.~\ref{fig:kag_m1_lines}b,
or parallel line defects (alternating
in direction), as in Fig.~\ref{fig:kag_m1_lines}a, allowing for 
entropy of order $L$.

\begin{table}
\begin{tabular}{llccccl}
\hline
\hline
Lattice & Complementary & $M$ & $\BB$ &
$\JJ$ & Ground\\
 & lattice & & & & state \\
\hline
pyrochlore & pyrochlore &
 $0$ & $+0.0136$ &  $-0.0016$ &  $\pi$-flux \\
pyrochlore & pyrochlore &
 $2$ &$-0.0091$ & $-0.00007$  & $0$-flux \\
checkerboard & square &
 $0$ &$-0.0649$ & $+0.0136$  & $0$-flux  \\
checkerboard & square &
 $2$ &$+0.0342$ & $-0.0091$  & $\pi$-flux  \\
capped kagom\'e & triangular &
 $0$ &$+0.0376$ & $+0.0052$  & $2 \times 2$  \\
kagom\'e & triangular &
 $1$ &$-0.0173$ & $-0.0037$  &$0$-flux\\
\hline
\hline
\end{tabular}
\caption{ \label{tab:summary}
Effective complementary lattice Ising Hamiltonian~(\ref{eq:comp_heff})
coefficients for the various models that we consider.
The coefficients were obtained by the expansion~(\ref{eq:h_expand}),
with $n\!\ge\!30$ and extrapolated to $n\!\to\!\infty$.
The complementary lattice ground state is given in the right column, and it
corresponds, in all cases, to a family of states in the direct lattice.}
\end{table}

\section{Conclusions}\label{sec:concl}

In this paper, we have presented a thorough semiclassical linear spin-wave
theory of the pyrochlore lattice antiferromagnetic Heisenberg model.
We found that for any collinear classical ground state, 
half of the  spin-wave modes are \emph{generic zero modes}, i.e. 
modes that have no dispersion, and that are the same for any state.
Of the remaining $N_s/2$ \emph{ordinary modes}, some are optical in nature,
while others vanish along the major axes in the Brillouin zone. 
These non-generic zero modes carry divergent fluctuations.
We find that the \emph{divergent modes} are closely related to
\emph{gauge-like} transformations that relate exactly divergent (to this order
in $1/S$) states.

In Sec.~\ref{sec:zeropoint}, we studied the (harmonic)
zero-point energy and find that
it can be expanded in loops, to obtain an
\emph{effective Hamiltonian}~(\ref{eq:heff}) written in terms
of \emph{flux variables}. 
We find the effective Hamiltonian coefficients using re-summation of the 
Taylor expansion, via a Bethe lattice approximation, in which we take advantage
of the fact that all state-independent terms can be evaluated to excellent
accuracy using closed paths on a coordination-$4$ Bethe lattice.
We have calculated the harmonic zero-point energy using numerical
diagonalization of the spin-wave spectrum to demonstrate 
that the effective Hamiltonian provides us in good agreement with exact results.

In Sec.~\ref{sec:noncoll}, we showed that any non-collinear deviation away from 
a collinear state results in an energy increase, and thus justified the search 
for a ground state among collinear states only.
While we did show that the collinear states are local energy minima in the space
of all classical ground states, we did not show that there are no other, 
non-collinear minima. 
More work would be required to rigorously prove this point.

It is instructive to compare our results to the well-known results on the
kagom\'e lattice.
In both cases, there is an extensive classical degeneracy, associated
with all states satisfying Eq.~(\ref{eq:cl_const}). 
Whereas in the pyrochlore collinear states are favored by the zero-point energy,
in the kagom\'e, which is composed of triangles, there are no collinear 
classical ground states.
It turns out from the harmonic spin-wave theory of the kagom\'e, that all 
\emph{coplanar} states satisfying~(\ref{eq:cl_const}) are exactly
degenerate~\cite{ritchey}, so that the entropy remains extensive.
Here, on the other hand, we find that, 
as suggested in Ref.~\onlinecite{clh_harmonic},
the pyrochlore harmonic order ground states 
are only a small subset of all of the collinear classical ground states,
and that the $\OO(L^3)$ classical entropy is
reduced to an entropy of, at most, $\OO(L \ln L)$ (and at least $\OO(L)$).

In order to search for a unique ground state one must go to higher order terms
in expansion~(\ref{eq:ham_exp})\cite{uh_quartic}, as has been done previously
for the kagom\'e lattice~\cite{chubukov+chan}.
In that case, anharmonic corrections to the zero-point energy resulted in a
selection of a unique kagom\'e ground state.
We expect that the spin-wave modes with divergent Gaussian fluctuations 
described in this paper will play a significant role in the anharmonic ground 
state selection.

In Sec.~\ref{sec:other_models} we showed that the loop expansion approach to the
effective Hamiltonian can be easily applied to the case of a magnetization 
plateau induced by a large magnetic field, as well as to other \emph{bisimplex}
lattices that support collinear spin configurations as classical ground states.
We applied this to find the effective Hamiltonian coefficients and ground state
of the $M\!=\!2$ pyrochlore, the $M\!=\!0$ and $M\!=\!2$ checkerboard model,
the $M\!=\!1$ kagom\'e model, and the \emph{capped kagom\'e}
model~\cite{capped_kagome}, which is
the only case that we have encountered where the ground state fluxes are
non-uniform.
Together with Prashant Sharma, we have recently~\cite{uh_LN}
employed a similar effective Hamiltonian on 
the large-$N$ Heisenberg model on the pyrochlore.
We believe that this type of approach can and should be used more often
in determining the ground state of highly frustrated models with many degrees
of freedom.

\begin{acknowledgments}
We would like to thank
Mark Kvale for access to his unpublished work and
Roderich Moessner, Oleg Tchernyshyov,
and Prashant Sharma for discussions.
This work was supported by NSF grant DMR-0240953.

\end{acknowledgments}

\appendix
\section{Bethe lattice paths} \label{app:bethe}
Consider a Bethe lattice of coordination $z$, and $N_B$ sites.
We shall assume that $N_B$ is infinite so that the translational 
symmetry of the lattice is conserved.
We would like to find the number of
paths of length $k$ that start and end at a particular site.
Since all sites are equivalent, the total number of paths would just be $N_B$
times this quantity.

Define the following values:
\begin{itemize}
\item{$f_k$: The number of paths of length $2k$ starting and ending at a 
particular site $\alpha$}.
\item{$g_k$: Same as $f_k$, but counting only paths that do not return to the
origin until the last step.}
\item{$\ft_k$: Same as $f_k$, but the origin $\alpha$ only has $z\!-\!1$
nearest neighbors.}
\item{$\gt_k$: Same as $g_k$, but the origin $\alpha$ only has $z\!-\!1$
nearest neighbors.}
\end{itemize}
The number $f_k$ can be calculated from $\{g_i: i\le k\}$. E.g.
\begin{equation}
f_3 = g_3 + 2 g_1 g_2 + g_1^3 \,.
\end{equation}
One way of calculating the coefficients of these expansions in by means of
generating functions
\begin{equation}
G=\sum_{k=1} g_k x^k \,, \qquad F=1+ G+G^2+\ldots=\frac{1}{1-G}\,,
\end{equation}
on the other hand 
\begin{equation}
F=\sum_k f_k  x^k \,, \qquad
f_k = \frac{1}{k!} \left.\frac{\partial^k F}{\partial x^k}\right|_{x=0} \,.
\end{equation}
$\ft_k$ can be calculated from $\{\gt_i: i\le k\}$ in an identical fashion.

We calculate $g_k$, $\gt_k$ and $\ft_k$ by induction:
\begin{itemize}
\item{$g_0=\gt_0=\ft_0=1$.}
\item{Find $g_k=z \ft_{k-1}$, $\gt_k=(z-1) \ft_{k-1}$.}
\item{Calculate $\ft_k$ from $\{\gt_i: i\le k\}$.}
\end{itemize}
Note that we always obtain $\ft_k \sim (z-1)^k$, $\gt_k \sim (z-1)^k$, and 
$g_k \sim z (z-1)^k$, and the prefactor depends only on $k$.
Once $\{g_k\}$ have been calculated we find
$\{f_k\}$, and calculate the Bethe lattice harmonic energy by
Eq.~(\ref{eq:expand}) (where the Bethe lattice takes the place of the $N_B$-site
simplex lattice)
\begin{eqnarray} \label{eq:ebethe}
E_\harm(\mathrm{Bethe}) &=&
S\sqrt{A} \sum_{n=0} C_n \sum_{k=0}^n  \frac{(-4)^{n-k}}{A^k} \comb{n}{k}
f_k N_B \nonumber \\
&&- SN_s \,,
\end{eqnarray}
or similarly for a non-zero magnetization plateau using Eq.~(\ref{eq:h_expand}).
$E_\harm(\mathrm{Bethe})$ is our approximation for the 
constant term $E_0$ in the energy.

\begin{figure}
\resizebox{8cm}{!}{\includegraphics{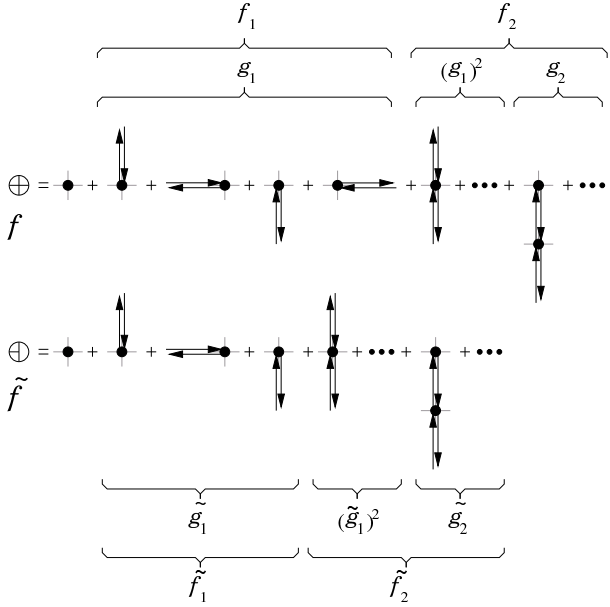}}
\caption{ \label{fig:diag}
\footnotesize Diagrammatic representation terms of summation of $f_k$ and
$\ft_k$, which is used in the loop calculation of Apps.~\ref{app:loop}
and~\ref{app:bethe}.
The difference between the two terms is that whereas $f_k$ is computed on 
a coordination $z$ Bethe lattice, in $ft_k$ the origin has only $z\!-\!1$ 
neighbors, and all other sites have $z$ neighbors.
}
\end{figure}

\section{Calculating loop coefficients} \label{app:loop}

\begin{figure}[t!]
\resizebox{!}{4cm}{\includegraphics{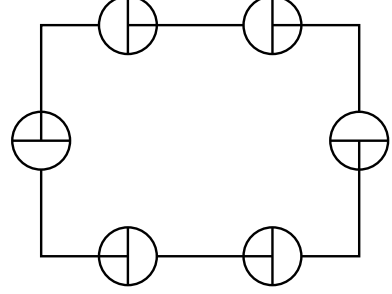}}
\caption{ \label{fig:hex_diag}
\footnotesize Diagrammatic representation of the paths 
included in calculation of the hexagon coefficient.
Each node along the loop is ``dressed'' by a Bethe lattice factor,
$\ft{k}$, as shown in Fig.~\ref{fig:diag}.}
\end{figure}

Here we provide the details of our calculation the coefficient $K_{2l}$ in 
Eq.~(\ref{eq:heff}).
Consider the terms in the expansion~(\ref{eq:expand}) that involve loops of
length $2l$ and no other loops.
In the Bethe lattice approximation, we assume that all of the paths
involving a loop, can be viewed as a decorated loops, i.e.,
a loop with self-retracing paths
(i.e. equivalent to Bethe lattice paths) emanating from each site.
This means that within our approximation, all loops of length $2l$ are
equivalent, and therefore, for any $k$, $\tr{\muvec^{2k}}$ in the expansion is
equal to $\Phi_{2l}$ multiplied by $f_k^l$, the number of decorated paths of
length $2k$ involving a particular loop of length $2l$.
The effective Hamiltonian~(\ref{eq:heff}) coefficient can therefore be written
\begin{equation} \label{eq:heff_coeff}
K_{2l} = 
S\sqrt{A} \sum_{n=0} C_n \sum_{k=0}^n  \frac{(-4)^{n-k}}{A^k} \comb{n}{k}
f_k^l\,.
\end{equation}
In the case of a large magnetic field of Sec.~\ref{sec:field},
Eq.~(\ref{eq:heff_coeff}) would be replaced by the appropriate expression
from Eq.~(\ref{eq:h_expand}).
\begin{eqnarray} \label{eq:h_heff_coeff}
K_{2l}(M) &=& S\sqrt{A} \sum_{n=0}^\infty
\frac{C_n}{A^n}\sum_{k=0}^n \sum_{j=0}^{n-k}  \tcomb{n}{k}{j} \nonumber \\ 
&& \times
 (M^2-4A)^{n-k-j} 
(-2M)^j f_k^l \,.
\end{eqnarray}
We want to count the number of paths of length $2k$ involving a particular
simple loop of length $2l$, and no other loops, such that
each site along the loop may be an origin of a self-retracing path 
(diagrammatically shown in Fig,~\ref{fig:hex_diag}).
As explained in Sec.~\ref{sec:loop}, in order to avoid double counting, we must 
consider trees whose origin has only $z-1$ nearest
neighbors. Fortunately, we have already calculated such terms in
App.~\ref{app:bethe}, i.e. the terms $\ft_i$.

All we have to do is to find all of the ways of distributing $k-l$ steps that 
are not part of the loop, among $2l$ sites,
and take the product of $\ft$ for each of those.
In more concrete terms, for a given $k$, the number of possible paths involving
a particular loop of length $2l$ is 
\begin{equation}
f^{l}_{k} \equiv 4k
\sum_{n=1}^{2l} \comb{2l}{n} 
\sum_{\sum i_j = k-l}
(k-l)! \frac{\ft_{i_1}}{i_1!}\frac{\ft_{i_2}}{i_2!}
\cdots \frac{\ft_{i_n}}{i_n!} \,.
\end{equation}
We have multiplied the sum by $4k$ because we can start anywhere along the path
and go in any of two directions.
Plugging the results, into Eq.~(\ref{eq:heff_coeff})
(or~(\ref{eq:h_heff_coeff})),
we obtain the effective Hamiltonian coefficients.

\begin{thebibliography}
\bibitem{footnote_1}{Alternatively, the lattice may be viewed as an FCC lattice
 of disjoint tetrahedra.}
\bibitem{footnote_2}{The energies in Table I of Ref.~\onlinecite{clh_harmonic}
are higher
by exactly $+S$ per spin compared to the results presented here, because that
calculation omits a constant term arising in the Holstein-Primakoff expansion.}
\end{thebibliography}
\end{document}